\documentclass[12pt]{article}

\usepackage{amsmath}
\usepackage{amssymb}
\usepackage{epsfig}
\usepackage{graphicx}
\usepackage{cite}

\newcommand{\capdef}{}
\newcommand{\mycaption}[2][\capdef]{\renewcommand{\capdef}{#2}%
        \caption[#1]{{\footnotesize #2}}}
\makeatletter
\renewcommand{\fnum@table}{\textbf{\tablename~\thetable}}
\renewcommand{\fnum@figure}{\textbf{\figurename~\thefigure}}
\makeatother

\newcounter{myenumi}

\renewcommand{\themyenumi}{\roman{myenumi}}
{\end{list}}

\newlength{\myem}
\newcounter{mysubequation}[equation]

\makeatletter
\renewcommand{\section}{\@startsection{section}{1}{0em}{-\baselineskip}%
{\baselineskip}{\normalfont\large\bfseries}}
\renewcommand{\subsection}%
{\@startsection{subsection}{2}{0em}{-0.7\baselineskip}%
{0.7\baselineskip}{\normalfont\bfseries}}
\makeatother

\newcommand{\bi}{\begin{itemize}}
\newcommand{\ei}{\end{itemize}}

\newcommand{\be}{\begin{equation}}
\newcommand{\ee}{\end{equation}}
\newcommand{\bea}{\begin{eqnarray}}
\newcommand{\eea}{\end{eqnarray}}

\newcommand{\ldm}{\Delta m_{31}^2}
\newcommand{\sdm}{\Delta m_{21}^2}
\newcommand{\deltacp}{\delta_{\mathrm{CP}}}
\newcommand{\stheta}{\sin^2(2 \theta_{13})}

\newcommand{\cp}{\emph{CP}}

\DeclareMathOperator{\im}{Im}
\DeclareMathOperator{\re}{Re}

\DeclareMathOperator{\diag}{diag}

\newcommand{\ie}{{\it i.e.}}

\newcommand{\eg}{{\it e.g.}}

\newcommand{\cf}{{\it cf.}}

\newcommand{\etc}{{\it etc.}}
\newcommand{\eq}{Eq.}
\newcommand{\eqs}{Eqs.}

\newcommand{\fig}{Fig.}

\newcommand{\Ref}{Ref.}
\newcommand{\Refs}{Refs.}
\newcommand{\Sec}{Sec.}
\newcommand{\Secs}{Secs.}

\newcommand{\Tab}{Table}

\newcommand{\ReactorI}{{\sc Reactor-I}}
\newcommand{\ReactorII}{{\sc Reactor-II}}

\newcommand{\NuFactII}{{\sc NuFact-II}}

\newcommand{\equ}[1]{\eq~(\ref{equ:#1})}
\newcommand{\figu}[1]{\fig~\ref{fig:#1}}

\begin{document}

\begin{titlepage}

\renewcommand{\thefootnote}{\alph{footnote}}



\renewcommand{\thefootnote}{\fnsymbol{footnote}}
\setcounter{footnote}{-1}

{\begin{center}
{\large\bf
Damping signatures in future neutrino oscillation
experiments
} \end{center}}
\renewcommand{\thefootnote}{\alph{footnote}}

\vspace*{.8cm}
\vspace*{.3cm}
{\begin{center} {\large{\sc
 		    Mattias~Blennow\footnote[1]{\makebox[1.cm]{Email:}
                mbl@theophys.kth.se},
                Tommy~Ohlsson\footnote[2]{\makebox[1.cm]{Email:}
                tommy@theophys.kth.se}, and
                Walter~Winter\footnote[3]{\makebox[1.cm]{Email:}
                winter@ias.edu}~
                }}
\end{center}}
\vspace*{0cm}
{\it
\begin{center}

\footnotemark[1]${}^,$\footnotemark[2]%
Division of Mathematical Physics, Department of
Physics, School of Engineering Sciences, Royal Institute of Technology
(KTH) -- AlbaNova University Center, \\ Roslagstullsbacken 11,
106~91~~Stockholm, Sweden

\footnotemark[3]%
       School of Natural Sciences, Institute for Advanced Study, \\
       Einstein Drive, Princeton, NJ 08540, USA

\end{center}}

\vspace*{1.5cm}

{\Large \bf
\begin{center} Abstract \end{center}  }

We discuss the phenomenology of damping signatures in the neutrino
oscillation probabilities, where either the oscillating terms or the
probabilities can be damped. This approach is a possibility for tests
of non-oscillation effects in future neutrino oscillation experiments,
where we mainly focus on reactor and long-baseline experiments. We
extensively motivate different damping signatures due to small
corrections by neutrino decoherence, neutrino decay, oscillations into
sterile neutrinos, or other mechanisms, and classify these signatures
according to their energy (spectral) dependencies. We demonstrate, at
the example of short baseline reactor experiments, that damping can
severely alter the interpretation of results, \eg, it could fake a
value of $\stheta$ smaller than the one provided by Nature. In
addition, we demonstrate how a neutrino factory could constrain
different damping models with emphasis on how these different models
could be distinguished, \ie, how easily the actual non-oscillation
effects could be identified. We find that the damping models cluster
in different categories, which can be much better distinguished from
each other than models within the same cluster.

\vspace*{.5cm}

\end{titlepage}

\newpage

\renewcommand{\thefootnote}{\arabic{footnote}}
\setcounter{footnote}{0}

\section{Introduction}

Neutrino oscillations are by far the most plausible description of
transitions among different neutrino flavor eigenstates
\cite{Fukuda:1998mi,Ahmad:2002jz,Ahmed:2003kj,Ahn:2002up,Eguchi:2002dm,Araki:2004mb,Ashie:2004mr}. However,
there have historically been other attempts in the literature to
describe these transitions with other mechanisms as well as neutrino
oscillations combined with such other mechanisms. These scenarios
include neutrino wave packet decoherence
\cite{Giunti:1998wq,Giunti:2003ax,Giunti:1992sx,Grimus:1998uh,Cardall:1999ze},
neutrino decay
\cite{Bahcall:1972my,Barger:1982vd,Valle:1983ua,Barger:1998xk,Pakvasa:1999ta,Barger:1999bg,Lindner:2001fx,Lindner:2001th},
oscillations into sterile neutrinos
\cite{Strumia:2002fw,Maltoni:2004ei}, neutrino absorption (see, \eg,
\Ref~\cite{DeRujula:1983ya}), and neutrino quantum decoherence
\cite{Lisi:2000zt,Benatti:2000ph,Adler:2000vf,Ohlsson:2000mj,Benatti:2001fa,Gago:2002na,Barenboim:2004wu,Barenboim:2004ev,Morgan:2004vv}. A
combined scenario is, for example, the combination of neutrino
oscillations and neutrino decay (see, \eg,
\Refs~\cite{Lindner:2001fx,Lindner:2001th}). Although these other
mechanisms, leading to ``non-standard effects'', are not such
successful descriptions for flavor transitions as neutrino
oscillations are (in fact, they are strongly disfavored
\cite{Ashie:2004mr,Araki:2004mb}), they could still give rise to small
corrections to the neutrino oscillations.  These non-standard effects
need to be described in a framework together with neutrino
oscillations and can be constrained by current and future experiments
(see, \eg, \Ref~\cite{Valle:2003uv} for a recent review). Thus, we
will assume that the leading order effect in neutrino flavor
transitions is due to neutrino oscillations, whereas the
next-to-leading order effects are described by different ``damping
mechanisms'' of the neutrino oscillations.

Since any non-standard effect may point towards new interesting
physics beyond the standard model, the test of small corrections due
to these effects should be one of the main objectives in future
high-precision neutrino oscillation physics. The assumption of
standard three-flavor neutrino oscillations will inevitably lead to an
erroneous derivation of the elements of the mixing matrix $U$ or the
mass squared differences. We therefore define ``non-oscillation
effects'' as any modification of the three-flavor neutrino oscillation
probabilities in vacuum as well as in matter. For example, the LSND
anomaly~\cite{Aguilar:2001ty} could be an indication of
non-oscillation effects according to this definition. Since future
reactor and long-baseline neutrino oscillation experiments are
expected to have high-precisions to the subleading neutrino
oscillation parameters $\stheta$ and $\deltacp$, we mainly discuss the
impact of non-oscillation effects or possible constraints on the
non-oscillation effects in the context of these experiments.

In principle, one could think of many different approaches to test
non-oscillation effects with future long-baseline experiments:
\begin{description}
\item[Neutral-currents] can be used to test the conservation of
probability, \ie, $P_{\alpha e} + P_{\alpha \mu} + P_{\alpha \tau} =1$ (see,
\eg, \Ref~\cite{Barger:2004db}). However, at long-baseline
experiments, uncertainties in the neutral-current cross-sections and the 
charged-current contamination lead to a
precision of only about
$10~\%-15~\%$~\cite{Barger:2004db}.  In addition, even if some
non-oscillation effects are found, there will be no information 
on the nature
of the effects, whereas effects conserving the overall probability cannot
be detected at all.
\item[The detection of $\boldsymbol{\nu_\tau}$] can complement the
information on $P_{\alpha e}$ and $P_{\alpha \mu}$ to test the
conservation of probability (see, \eg, \Ref~\cite{Donini:2002rm}).
Since $\nu_\tau$ detection is much more sophisticated and less
efficient than the detection of $\nu_e$ and $\nu_\mu$ due to the higher
$\tau$ production thres\-hold, this is also a non-trivial test. If
there are non-oscillation effects, then the information will be 
better than in
the preceding case, since one will know which neutrino oscillation
probabilities are affected.
\item[Unitarity triangles] for the lepton sector can be
constructed~\cite{Farzan:2002ct,Zhang:2004hf}. However, since there is
no simple relationship between the quantities of the unitarity
triangles and the neutrino oscillation observables, this approach may
not be the most feasible for the lepton sector.
\item[Tests of distinctive signatures,] \ie, spectral (energy)
dependent effects, could directly identify certain classes of
non-oscillation effects
\cite{Valle:2003uv,Huber:2001zw,Huber:2001de,Huber:2002bi}. The
advantage of such tests is that the effect could be directly
identified if it produces a unique signature in the energy
spectrum. In addition, this test does not depend upon normalization
errors of the event rates, which are likely to constrain the first two
measurements. However, there might be strong correlations with the
neutrino oscillation parameters.
\end{description}

In addition, in the future, it may be possible to resolve the line
width and shape of the ${}^7$Be solar neutrino line
\cite{Bahcall:1993ej,Bahcall:1994cf} and extract the temperature
distribution as well as the modulation of this line, which could be
caused by next-to-leading order effects. Thus, performing very high-energy
resolution measurements of the ${}^7$Be line may be an idea how to
determine these next-to-leading order effects. Such possible precision
neutrino experiments include, for example, a bromine cryogenic thermal
detector proposed in \Refs~\cite{Fiorini:1991,Alessandrello:1995ih}.

In this study, we will focus on the tests of distinctive signatures in
which we introduce ``damping signatures'' as an abstract concept for a
class of possible effects entering at probability
level.\footnote{Although it will be possible to describe some of our
effects on Hamiltonian level, the Hamiltonian will not be Hermitian
anymore.} In general, small Hamiltonian effects, see, \eg,
\Ref~\cite{Benatti:2001fa}, may be as important as the kind of damping
effects that we will describe in this study. Such Hamiltonian effects
could lead to direct changes in the effective neutrino oscillation
parameters. Nevertheless, those effects cannot be treated in the
framework presented here. We will use the observation that
mechanisms, such as decoherence or decay, lead to exponential damping
in the neutrino oscillation probabilities. However, the effect might
be stronger for low or high energies, \ie, the spectral (energy)
dependence for the damping might be different.  A common feature for
many of the discussed models is that they will lead to less neutrinos
(of all active flavors) being detected than what is expected from the
three-flavor neutrino oscillations. For all other models, only the
oscillating terms of the neutrino oscillation probabilities will be
damped, while the total number of active neutrinos remains constant.
Note that the damping signature approach does not cover all possible
models, but many models can, at least in the limit of small
corrections, lead to some exponential damping effect.

Our study is organized as follows. In \Sec~\ref{sec:phenomenology}, we
will present and classify different forms of the damping
signatures. For the reader who is not interested in different models
for damping signatures, at least \Sec~\ref{sec:gendescription} and the
examples in \Tab~\ref{tab:models} should be read to be able to follow
the rest of the study. Next, in \Sec~\ref{sec:dampedprob}, we will
give and discuss the damped neutrino oscillation probabilities arising
from the effects described by their signatures. For the reader, who is
most interested in possible experimental implications,
\Sec~\ref{sec:dampedtwoflavor} should summarize the most relevant
features, whereas the rest of this section deals with the more
technical three-flavor cases. Then, in \Secs~\ref{sec:appl1} and
\ref{sec:appl2}, we will discuss the physics of these damping
signatures and give two different applications in the framework of a
complete experiment simulation. Especially, in \Sec~\ref{sec:appl1},
we demonstrate how such damping signatures can modify the
interpretation of physical results for future reactor experiments,
whereas in \Sec~\ref{sec:appl2}, we discuss how a neutrino factory
could constrain different damping signatures and how these different
signatures could be distinguished. Finally, in \Sec~\ref{sec:summary},
we will summarize our work and present our conclusions.

\section{Phenomenology of damping signatures}
\label{sec:phenomenology}

In this section, we motivate, in a phenomenological manner, the form
of the damping signatures used for the rest of this study.

\subsection{General description of damped neutrino oscillations in vacuum}
\label{sec:gendescription}

We start with three-flavor neutrino oscillations in vacuum, which can
be described by the (undamped) vacuum oscillation probabilities
\begin{eqnarray}
P_{\alpha \beta} \equiv  P(\nu_\alpha \rightarrow \nu_\beta) & = & \left| \langle \nu_\beta |
U \, \diag \left( 1 ,  \exp \left( -{\rm i}\frac{\sdm L}{2 E} \right),
 \exp \left( -{\rm i}\frac{\ldm L}{2 E} \right)
\right) \, U^\dagger | \nu_\alpha \rangle \right|^2 \nonumber \\ & = &
\sum\limits_{i,j=1}^{3} U_{\alpha j} \, U_{\beta j}^* \, U_{\alpha
i}^* \, U_{\beta i} \, \exp(- {\rm i} \Phi_{ij} ).
\end{eqnarray}
Here $U$ is the leptonic mixing matrix in vacuum, $\Delta m_{ij}^2
\equiv m_i^2 - m_j^2$ the mass squared difference, and $\Phi_{ij}
\equiv \Delta m_{ij}^2 L/(2 E)$ the oscillation phase. By defining
$$
J_{ij}^{\alpha\beta} \equiv U_{\alpha j}
U_{\beta j}^*  U_{\alpha i}^* U_{\beta i}
\quad {\rm and} \quad \Delta_{ij} \equiv \frac{\Delta
m_{ij}^2L}{4E} \equiv \frac{m_i^2-m_j^2}{4E}L = \frac{\Phi_{ij}}2,
$$
the oscillation probabilities may be written as
\begin{eqnarray}
P_{\alpha\beta} &=&
\sum_{i,j = 1}^3 \re(J_{ij}^{\alpha\beta}) -
4 \sum_{1\leq i<j \leq 3} \re(J_{ij}^{\alpha\beta})\sin^2 (\Delta_{ij}) -
2 \sum_{1\leq i<j \leq 3} \im(J_{ij}^{\alpha\beta})\sin (2\Delta_{ij}) 
\nonumber \\
\label{equ:vacprob}
&=&\sum_{i=1}^3 J_{ii}^{\alpha\beta} +
2 \sum_{1\leq i < j \leq 3}
|J_{ij}^{\alpha\beta}| \cos(2\Delta_{ij}+\arg J_{ij}^{\alpha\beta}),
\end{eqnarray}
where, in the first line of the equation, the first two terms are
\cp-conserving and the third term is the source of any \cp~violation,
this corresponds to $\arg J_{ij}^{\alpha\beta}$ being the source of
any \cp~violation in the second line. As will be discussed, there may
be reasons to assume that \equ{vacprob} does not give the correct
neutrino oscillation probabilities. Effects that might spoil this
approach of calculating neutrino oscillations probabilities include
loss of wave packet coherence and neutrino decay. The effective result
of such processes is to introduce damping factors to the oscillating
terms of the neutrino oscillation probabilities. We define a general
damping effect to be an effect that alters the neutrino oscillation
probabilities to the form
\begin{eqnarray}
P_{\alpha\beta} &=& \sum\limits_{i,j=1}^{3}
U_{\alpha j} \, U_{\beta j}^* \, U_{\alpha
i}^* \, U_{\beta i} \, \exp(- {\rm i} \Phi_{ij} ) D_{ij} \nonumber \\
&=&
\sum_{i,j = 1}^3 \re(J_{ij}^{\alpha\beta})D_{ij} -
4 \sum_{1\leq i<j \leq 3} \re(J_{ij}^{\alpha\beta})D_{ij}\sin^2 (\Delta_{ij}) -
2 \sum_{1\leq i<j \leq 3} \im(J_{ij}^{\alpha\beta})D_{ij}\sin (2\Delta_{ij})
\nonumber \\
&=&\sum_{i=1}^3 J_{ii}^{\alpha\beta} D_{ii} +
2 \sum_{1\leq i < j \leq 3}
|J_{ij}^{\alpha\beta}| D_{ij}\cos(2\Delta_{ij}+\arg J_{ij}^{\alpha\beta}),
\label{equ:damping}
\end{eqnarray}
where the damping factors
\begin{equation}
\label{equ:dfactor}
D_{ij} = \exp\left(-\alpha_{ij}\frac{|\Delta m_{ij}^2|^\xi L^\beta}
{E^\gamma}\right)
\end{equation}
have been introduced and we have assumed that $D_{ij} =
D_{ji}$. Obviously, as $D_{ij} \rightarrow 1$, we regain the undamped
oscillation probabilities given in \equ{vacprob}. In \equ{dfactor},
$\alpha_{ij} \ge 0$ is a non-negative damping coefficient matrix, and
$\beta$, $\gamma$, and $\xi$ are numbers that describe the
``signature'', \ie, the $L$ ($\beta$) and $E$ ($\gamma$) dependencies
as well as the dependence on the mass squared differences. In
addition, the parameter $\xi$ implies two interesting cases:
\begin{description}
\item[$\boldsymbol{\xi>0}$:] In this case, only the oscillating terms
will be damped, since $\Delta m_{ii}^2 = 0$ by definition.
\item[$\boldsymbol{\xi=0}$:] The whole oscillation probability can be
damped (depending on $\alpha_{ij}$), since also the terms which are
independent of the oscillation phases are affected.
\end{description}
Therefore, we expect two completely different results for these two
cases.  In general, \equ{dfactor} introduces twelve new parameters,
which can be used to model many non-standard contributions that enter
on the oscillation probability (not Hamiltonian) level. We will give
some examples of such contributions below.  Although we expect these
contributions to be small, it is rather impractical to deal with that
many new parameters, which means that some simplifications need to be
made.  First of all, note that the parameter $\beta$ is not measurable
if only one baseline is considered and can therefore be absorbed in
$\alpha_{ij}$. For two baselines, it can, in principle, be resolved if
all the other parameters are known. Second, for a specific model,
there may be relations among different $\alpha_{ij}$'s that actually
imply much fewer independent parameters. For a very simple model, the
number of parameters can even reduce to one. Since we are mainly
interested in the spectral signatures, \ie, $\gamma$, we will often
use $\alpha_{ij} \equiv \alpha$ to estimate the magnitude of different
effects.  Third, it will turn out that the parameter $\xi$ is strongly
dependent on the model, since, as discussed above, it describes two
completely different classes of models.  Hence, we will finally end up
with one free parameter $\alpha$ and several fixed model dependent
parameters $\beta$, $\gamma$, and $\xi$.

\subsection{A model for damped neutrino oscillations in matter}

In some cases, we will use neutrino propagation in matter, since, for
instance, neutrino factories operate at very long-baselines for which
matter effects become important. We use an approach similar to
\equ{damping}, which should describe the damping signatures as minor
perturbations to neutrino oscillations in (constant) matter as long
as they are small enough:
\begin{equation}
P_{\alpha\beta} = \sum_{i,j = 1}^3 \re(\tilde J_{ij}^{\alpha\beta})\tilde D_{ij} -
4 \sum_{1\leq i<j \leq 3}
\re(\tilde J_{ij}^{\alpha\beta})\tilde D_{ij}\sin^2 (\tilde \Delta_{ij}) -
2 \sum_{1\leq i<j \leq 3}
\im(\tilde J_{ij}^{\alpha\beta})\tilde D_{ij}\sin (2\tilde\Delta_{ij}),
\label{equ:dampingmatter}
\end{equation}
where the tildes denote the effective parameters for neutrinos
propagating in matter (for instance, $\tilde J_{ij}^{\alpha\beta} =
\tilde{U}_{\alpha j} \, \tilde{U}_{\beta j}^* \, \tilde{U}_{\alpha
i}^* \, \tilde{U}_{\beta i}$, where $\tilde U$ is the effective
leptonic mixing matrix in matter, \ie, the matrix re-diagonalizing the
Hamiltonian with the matter potential included). In general, the
damping effects may not enter directly as multiplicative factors in
the interference terms among different matter
eigenstates.\footnote{For instance, some effect on Hamiltonian level,
such as neutrino absorption, would require a full re-diagonalization
of the effective Hamiltonian with the absorption terms included, see
the section ``Neutrino absorption'' below.} However, in this study, we
assume small damping effects that should act as perturbations which,
to leading order, give rise to neutrino oscillation probabilities in
matter of the same form as the ones in vacuum.

Thus, we use the propagation in constant matter and apply the damping
signatures to the mass eigenstates in matter. This means that we
discuss signatures which depend on the mass eigenstates in
matter. They may come from wave packet decoherence, neutrino decay,
neutrino oscillations into sterile neutrinos, neutrino absorption,
quantum decoherence, or other mechanisms. Strictly speaking, this
model does not describe many of these mechanisms exactly, since a
complete re-diagonalization of the Hamiltonian might be necessary
(such as for Majoron decay in matter; see, \eg,
\Refs~\cite{Berezhiani:1987gf,Giunti:1992sy}). However, we treat only
small effects in matter acting as a perturbation to the neutrino
oscillation mechanism and do not consider transitions from active into
active neutrinos, which would require a more complicated treatment
(such as decay into other active neutrino states). Therefore, this
model should be sufficient as a first approximation, since we will
later on use either short baselines or mainly discuss effects in the
$P_{\mu \mu}$ channel, which are not affected by matter effects to
first order in the ratio of the mass squared differences $\Delta
m_{21}^2/\Delta m_{31}^2$ and the mixing parameter $s_{13} \equiv
\sin(\theta_{13})$ \cite{Akhmedov:2004ny}.

\subsection{Examples of different damping signatures}
\label{sec:examples}

\begin{table}[t]
\begin{center}
\begin{tabular}{p{3.0cm}ccccc}
\hline
 \begin{tabular}{l} Damping type \end{tabular} & Signature $D_{ij}$ & Unit for $\alpha$ & $\beta$ & $\gamma$ & $\xi$ \\
\hline
\begin{tabular}{l} Wave packet \\ decoherence \end{tabular} &
$\exp \left( - \sigma_E^2 \frac{(\Delta m_{ij}^2)^2 L^2}{8E^4} \right)$ &
$\mathrm{MeV}^2$ or $\mathrm{GeV}^2$ & 2 & 4 & 2 \\
 \begin{tabular}{l} Decay \end{tabular} & $\exp \left( - \alpha \frac{L}{E} \right)$ &
$\mathrm{GeV \cdot km^{-1}}$ & 1 & 1 & 0 \\
 \begin{tabular}{l} Oscillations to $\nu_s$ \end{tabular}& $ \exp \left( - \epsilon \frac{L^2}{(2E)^2} \right)$ &
 $\mathrm{eV}^4$ & 2 & 2 & 0 \\
 \begin{tabular}{l} Absorption \end{tabular} & $\exp \left( - \alpha L E \right)$ &
$\mathrm{GeV}^{-1} \cdot \mathrm{km}^{-1}$   & 1 & $-1$ & 0 \\
 \begin{tabular}{l} Quantum \\ decoherence I \end{tabular} &  $\exp \left( - \alpha L E^2 \right)$ & $\mathrm{GeV}^{-2} \cdot \mathrm{km}^{-1}$ & 1 & $-2$ & 0 \\
\begin{tabular}{l} Quantum \\ decoherence II \end{tabular} &
 $\exp \left( - \kappa \frac{(\Delta m_{ij}^2)^2}{E^2} \right)$ &
$\mathrm{eV}^{-2} $ & 1 or 2 & 2 & 2 \\
\hline
\end{tabular}
\end{center}
\mycaption{\label{tab:models} Different examples for damping
signatures considered in this study. The parameter $\gamma$ represents
the spectral (energy) dependence of the signature. The parameter
$\alpha$ has in some places been re-defined for convenience (see main
text) unless it corresponds exactly to our definition of $\alpha$. The
quantum decoherence models I and II are two examples of signatures
motivated by quantum decoherence (see
\Tab~\ref{tab:qdecoherence}). The quantum decoherence model II absorbs
$\beta$ in the definition of $\kappa \equiv \alpha L^\beta$ in order
to describe two of the models from \Tab~\ref{tab:qdecoherence}. Note
that another commonly used quantum decoherence signature is the same
as the decay signature.}
\end{table}

The general damping signature in \equ{dfactor} seems to be very
abstract. Therefore, let us now give some motivations for such damping
signatures by different mechanisms, which are summarized in
\Tab~\ref{tab:models}.

\subsubsection*{Intrinsic wave packet decoherence}

Intrinsic wave packet decoherence is an effect that appears even in
standard neutrino oscillation
treatments~\cite{Giunti:1998wq,Giunti:2003ax,Giunti:1992sx,Grimus:1998uh,Cardall:1999ze}.
It naturally emerges from any quantum mechanical model that does not
assume neutrino mass eigenstates propagating as plane waves or from
any quantum field theoretical treatment.  In principle, intrinsic
decoherence may not be distinguishable from a macroscopic energy
averaging (see, \eg, discussions in
\Refs~\cite{Kiers:1996zj,Giunti:2003mv,Lipkin:2003st}). Therefore, it
is natural to expect that the test of this signature could be limited by the
knowledge on the energy resolution of the detector.

We adopt the treatment in \Ref~\cite{Giunti:1998wq}, which uses
averaging over Gaussian wave packets. In this approach, the loss of
coherence can only be described at probability level.  It leads to
factors $\exp \left[-(L/L_{ij}^{\mathrm{coh}})^2 \right]$ in
\equ{dfactor}, where $L_{ij}^{\mathrm{coh}} = 4 \sqrt{2} \sigma_x
E^2/|\Delta m_{ij}^2|$ and $\sigma_x$ is the spatial wave packet
width. In this case, the damping descriptions in vacuum and matter
using \eqs~(\ref{equ:damping}),~(\ref{equ:dfactor}), and~(\ref{equ:dampingmatter}) are
accurate. For the damping signature, we obtain
\begin{equation}
D_{ij} = \exp \left[ - \left( \frac{L}{L_{ij}^{\mathrm{coh}}}
\right)^2 \right] =
 \exp \left[  - \left( \frac{\sqrt{2}\sigma_E}{E} \frac{\Delta m_{ij}^2 L}{4 E} \right)^2 \right] =
 \exp \left(  - \sigma_E^2  \frac{(\Delta m_{ij}^2)^2 L^2}{8 E^4} \right)
 \label{equ:coherence}
\end{equation}
in vacuum and the analogous signature $\tilde{D}$ in matter. Here we
have introduced a wave packet spread in energy $\sigma_E \equiv 1/(2
\sigma_x)$, since we later will derive an upper bound for this
quantity and directly compare it to the energy resolution of a
detector. The typical units of $\sigma_E$ will be $\mathrm{MeV}$ or
$\mathrm{GeV}$.  By comparing \eqs~(\ref{equ:dfactor}) and
(\ref{equ:coherence}), we can identify $\alpha_{ij} = \sigma_E^2/8$,
$\beta=2$, $\gamma=4$, and $\xi=2$. Note that, in this case, the
$\alpha_{ij}$'s do not depend on the indices $i$ and $j$.

In order to better understand \equ{coherence}, we note that
$\Delta m_{ij}^2 L/(4 E)$ is of order unity for the first oscillation maximum:
\begin{equation}
D_{ij} = \exp \left[  - \left( \frac{\sqrt{2}\sigma_E}{E} \frac{\Delta m_{ij}^2 L}{4 E} \right)^2 \right] = \exp \left[  - \left( \frac{\sigma_E}{\sqrt{2} E} \Phi_{ij} \right)^2 \right] \simeq
\underbrace{
 \exp \left[  - \left( \frac{1}{\sqrt{2}\sigma_xE} \mathcal{O}(1)
   \right)^2 \right]}_{\mathrm{value \, at \, oscillation \, maximum}} \, .
 \label{equ:coherence2}
\end{equation}
{}From \equ{coherence2}, we find three major implications: First, it
means that no effect will be observed if $\sigma_E \ll E$, because the
oscillation phase is usually of order unity (or less).  Second, since
the decoherence damping factor always comes together with an
oscillation phase factor with the same $\Delta m_{ij}^2$ [\cf,
\equ{damping}], it will equally damp the solar and atmospheric
oscillating terms in one probability formula. This means for the
atmospheric oscillation experiments that if the solar contribution
cannot be neglected, its damping factor can also not be
neglected. Third, one expects the largest suppression for low energies
independent of the type of oscillation experiment (solar or
atmospheric), since in either case the experiment will be operated
close to the oscillation maximum.
Eventually, it is important to keep in mind that this decoherence
signature is not an intrinsic property of the neutrinos, but an effect
related to the production and detection processes. Therefore, the
parameter $\sigma_E$ could be different for different classes of
experiments.

\subsubsection*{Invisible neutrino decay}

Another example of a damping signature is neutrino decay (see, \eg,
\Refs~\cite{Bahcall:1972my,Barger:1982vd,Valle:1983ua,Barger:1998xk,Pakvasa:1999ta,Barger:1999bg}).
In particular, invisible decay, \ie, decay into particles invisible
for the detector, leads to a loss of three-flavor unitarity. In this
case, the neutrino evolution is given by an effective Hamiltonian
\begin{equation}
H_{\rm eff} = H - {\rm i}\Gamma,
\end{equation}
where $\Gamma \equiv \diag(a_1,a_2,a_3)/2$ in the neutrino mass eigenstate
basis, $a_i \equiv \Gamma_i/\gamma_i$, $\Gamma_i$ is the inverse
life-time of a neutrino of mass eigenstate $i$ in its own rest frame,
and $\gamma_i \equiv E/m_i$ is the time dilation factor. We note that
$H$ and $\Gamma$ are both diagonal in the neutrino mass eigenstate
basis. The neutrino oscillation probabilities may now be calculated as
usual with the exception that, in addition to the phase factor
$\exp[-{\rm i}m_i^2L/(2E)]$, a factor of $\exp[-\Gamma_i m_i L/(2 E)]$ is
obtained when evolving the neutrino mass eigenstate $\nu_i$. The
resulting neutrino oscillation probabilities are of the form of
\equ{damping} with
\begin{equation}
\label{equ:decay}
D_{ij} = \exp\left(-\frac{\alpha_i + \alpha_j}{2E} L\right),
\end{equation}
where $\alpha_i = \Gamma_i m_i$, in accordance with
\Refs~\cite{Lindner:2001fx,Lindner:2001th}. Thus, for neutrino decay,
the characteristic signature is $\alpha_{ij} = (\alpha_i +
\alpha_j)/2$, $\beta = \gamma = 1$, and $\xi = 0$.

An example of the above decay is Majoron decay into lighter sterile
neutrinos. In this case, it is plausible to assume a quasi-degenerate
neutrino mass scheme for the active neutrinos with approximately equal
decay rates for all mass eigenstates, since the decay products all
have to be considerably lighter than the active neutrinos to obtain
fast decay rates due to phase space. The decay rates of the
$\alpha_i$'s will then be approximately equal ($\alpha_i = \alpha$ for
all $i$) and will typically be given in units of
$\mathrm{GeV}/\mathrm{km}$. Note that the decay rate is an intrinsic
neutrino property, not an experiment-dependent quantity such as the
wave packet decoherence. We identify by the comparison of \equ{decay}
with \equ{dfactor} that $\alpha$ is the same quantity\footnote{In
general, we do not change the symbol for $\alpha$ if its is exactly
the same as the one in \equ{dfactor}.  However, if there are
additional factors absorbed in $\alpha$, then we re-define the name
(such as for wave packet decoherence).}, $\beta=\gamma=1$, and
$\xi=0$. In matter, we use the analogous signature, \ie, we let the
mass eigenstates in matter decay.  In general, this is only a first
approximation, since, for example for Majoron decay in matter, a
re-diagonalization of the complete Hamiltonian may be necessary; see,
\eg, \Refs~\cite{Berezhiani:1987gf,Giunti:1992sy}.  However, as we
have assumed equal decay rates for all eigenstates, it should describe
the problem exactly, since the mass eigenstates in matter will also
decay with equal rates. In different decay models, the $\alpha_{ij}$'s
may not be identical anymore. For example, for a hierarchical mass
scheme with a normal hierarchy, the mass eigenstate $m_3$ decays much
faster than the other two. In this case, the observed effects in
atmospheric oscillations would qualitatively be similar, but about a
factor of two smaller (since mainly $m_2$ and $m_3$ participate in the
oscillation and only one of them decays). However, in matter such a
model is much more difficult to treat, since it is not easy to
identify the mass eigenstate in matter after the diagonalization of
the Hamiltonian. This problem does not occur with equal decay rates.

\subsubsection*{Oscillations into sterile neutrinos}
\label{sec:oscsteriles}

A natural description for the LSND result~\cite{Aguilar:2001ty} is a
light sterile neutrino (\ie, not a weakly interacting neutrino) that
is mixing with the active neutrinos.  This description is now
disfavored for the LSND
experiment~\cite{Strumia:2002fw,Maltoni:2004ei}, but small admixtures
of light sterile neutrinos cannot be entirely excluded. In particular
for slow enough oscillations into sterile neutrinos, the oscillation
signature $\sin^2 \Delta_{4i}$ with $\Delta_{ij} \equiv \Delta
m_{ij}^2 L/(4E)$ translates into damping signatures:
\begin{equation}
 1 - \epsilon \, \sin^2 \left(\frac{\Delta m_{4i}^2 L}{4E} \right)
\simeq 1 - \epsilon \left( \frac{\Delta m_{4i}^2 L}{4 E} \right)^2
\simeq \exp \left[
-  \epsilon \left( \frac{\Delta m_{4i}^2 L}{4 E} \right)^2 \right] \, ,
\end{equation}
where $\epsilon$ represents the magnitude of the mixing. Thus, the
damping coefficient $\alpha$ will (in this case) be determined by the
sizes of the mixing and the mass squared differences $\Delta
m_{4i}^2$.  We use as a model in vacuum (and the same form in matter)
\begin{equation}
D_{ij} = \exp \left( - \alpha_{ij} \frac{L^2}{(2E)^2} \right) = \exp \left( - \epsilon \frac{L^2}{(2E)^2} \right) \, ,
\label{equ:oscillations}
\end{equation}
where $\epsilon$ contains the information on mixing and $\Delta m^2$
and will be given in units of $\mathrm{eV}^4$ (the mixing factor is
dimensionless). Thus, we identify by comparison of \equ{oscillations}
with \equ{dfactor} that $\alpha_{ij} = \epsilon/4$, $\beta=\gamma=2$,
and $\xi=0$. Note that we only discuss effects independent of $i$ and
$j$, which simplifies the problem, but restricts the number of
applications tremendously. In addition, although the coefficient
$\epsilon$ is not experiment dependent (since it is an intrinsic
neutrino property here), it may (partly because of the independence on
$i$ and $j$) depend on the oscillation channel and mass scheme. As an
example, let us consider $P_{\mu\mu}$ and a mass scheme with $\Delta
m_{21}^2 \ll \Delta m_{43}^2 < \Delta m_{31}^2$, \ie, $\ldm$ is the
largest mass squared difference.  In this case, one can show that to
first approximation $\epsilon \simeq U_{\mu 4}^2 \, U_{\mu 3}^2 (
\Delta m_{43}^2)^2$ (for \cp~conservation). Thus, $\epsilon$ is
suppressed by the flavor content of $\nu_4$ in $\nu_\mu$ and the extra
mass squared difference, since all the other mass squared differences
with the sterile state are absorbed into the atmospheric oscillation
terms. In general, it should be noted that sterile neutrinos are
not affected in the same way as active neutrinos when propagating
through matter (\ie, there is a phase difference due to the
neutral-current interactions between matter and the active neutrino
flavors). However, the exponential damping signature for oscillations
into sterile neutrinos presented here is only valid for short
baselines, where matter effects have not yet developed.

\subsubsection*{Neutrino absorption}

When neutrinos propagate through matter, there is a small chance of
absorption. Neutrino absorption can be described in a fashion similar
to neutrino decay. In this case, we assume that an effective
Hamiltonian is given by
\begin{equation}
H_{\rm eff} = H - {\rm i}\Gamma,
\end{equation}
where $H$ is the usual neutrino Hamiltonian in matter, $\Gamma$ is
given by
\begin{equation}
\Gamma = \rho \diag(\sigma_e, \sigma_\mu, \sigma_\tau)/2
\end{equation}
in the flavor eigenstate basis, $\rho$ is the matter density, and
$\sigma_\alpha$ is the absorption cross-section for a neutrino of
flavor $\alpha$. If we assume the cross-sections to be relatively
small, then the eigenstates of $H_{\rm eff}$ will not differ
significantly from the orthogonal eigenstates of $H$. Thus, the first
order corrections to the eigenvalues of the effective Hamiltonian will
be
\begin{equation}
\delta E_i^{(1)} = -{\rm i} \Gamma_{ii} =
-{\rm i} \frac \rho 2 \sum_\alpha |U_{\alpha i}|^2 \sigma_\alpha
\equiv
-{\rm i}\frac \rho 2 \sigma_i,
\end{equation}
where $\sigma_i$ is an effective cross-section for a neutrino of mass
eigenstate $i$. The neutrino oscillation probability is now given by
an expression of the form of \equ{damping} with
\begin{equation}
D_{ij} = \exp\left(
-\frac{\sigma_i + \sigma_j}{2} \rho L
\right)= \exp\left(
-\frac{\sigma_i(E) + \sigma_j(E)}{2} \rho L
\right) \, ,
\end{equation}
where we have assumed a constant matter density $\rho$. The signature
of this scenario is given by $\beta = 1$ and $\gamma$ is equal to
minus the power of the energy dependence of the cross-sections. It
should be observed that, since the cross-sections increase with
energy, $\gamma$ will be a negative number.

If all neutrino flavor cross-sections were equal (or approximately
equal), then the effective matter eigenstate cross-sections would also
be equal.\footnote{Because of the higher $\tau$ production threshold,
the $\nu_e$ and $\nu_\mu$ cross-sections are in fact considerably
larger than the $\nu_\tau$ cross-section
\cite{Dutta:1999jg,Paschos:2001np,Kretzer:2002fr}. However, for these
low energies the standard absorption effects are anyway small.} For
the neutrino energies relevant to a neutrino factory, the
neutrino-nucleon cross-sections are approximately linear in energy
\cite{Gandhi:1998ri}. Thus, in this energy range, the damping
signature is given by $\alpha = \rho \sigma(E_0)/E_0$, $\beta = 1$,
$\gamma = -1$, and $\xi = 0$, where $\sigma(E_0)$ is the cross-section
at energy $E_0$. At higher energies, the cross-sections increase at a
slower rate and if damping effects are studied at these energies, then
the effective damping parameter $\gamma$ lies in the interval $-1 <
\gamma < 0$.

It should be noted that the standard neutrino absorption effects
(by weak interactions) are
very small for energies typical for neutrino oscillation experiments. However,
there could be non-standard absorption effects and the cross-sections
of these effects should behave in a manner similar to the standard
absorption.

\subsubsection*{Quantum decoherence}

It has been argued that quantum decoherence could be an alternative
description of neutrino flavor transitions. Fits to data by different
collaborations (\eg, Super-Kamiokande~\cite{Ashie:2004mr} and
KamLAND~\cite{Araki:2004mb}) have been performed and these clearly
disfavor a decoherence explanation for neutrino
flavor transitions. However, quantum decoherence may still be a
marginal effect in addition to neutrino oscillations and could give
rise to damping factors of the type given in \equ{dfactor}.

Quantum decoherence arises when a neutrino system is coupled to an
environment (or a reservoir or a bath), which could consist of, for
example, a space-time ``foam''~\cite{Lisi:2000zt} leading to new
physics beyond the standard model. Thus, quantum decoherence may be a
feature of quantum gravity. In order to find the formulas describing
quantum decoherence, it is necessary to use the Liouville
equation with decoherence effects of the Lindblad form~\cite{Lindblad:1975ef}.

Throughout the literature
\cite{Lisi:2000zt,Benatti:2000ph,Adler:2000vf,Gago:2000qc,Gago:2000nv,Ohlsson:2000mj,Benatti:2001fa,Gago:2002na,Barenboim:2004wu,Barenboim:2004ev,Morgan:2004vv},
the effects of loss of quantum coherence in neutrino oscillations have
been studied. Although the signatures derived by different authors
seem to vary, the decoherence effects are of the same form as
\equ{dfactor}. However, there might be additional effects on the
oscillation phases. In \Tab~\ref{tab:qdecoherence}, we give a brief
summary of some of the signatures that are present in the literature,
these examples could be used to motivate the numerical testing of such
signatures.
\begin{table}
\begin{center}
\begin{tabular}{lccccc}
\hline
Reference & Signature $D_{ij}$ & Unit for $\alpha$ & $\beta$ & $\gamma$ & $\xi$  \\
\hline
Lisi {\it et al.} \cite{Lisi:2000zt} and
Morgan {\it et al.} \cite{Morgan:2004vv} & $\exp \left( - \alpha L \right)$ &
$\mathrm{km}^{-1}$ & 1 & 0 & 0 \\
Lisi {\it et al.} \cite{Lisi:2000zt} and
Morgan {\it et al.} \cite{Morgan:2004vv} &  $\exp \left( - \alpha \frac{L}{E} \right)$ &
$\mathrm{GeV} \cdot \mathrm{km}^{-1}$
 & 1 & 1 & 0  \\
Lisi {\it et al.} \cite{Lisi:2000zt} and
Morgan {\it et al.} \cite{Morgan:2004vv} & $\exp \left( - \alpha L E^2 \right)$ &
$\mathrm{GeV}^{-2} \cdot \mathrm{km}^{-1}$ & 1 & $-2$ & 0  \\
 Adler \cite{Adler:2000vf} & $\exp \left( - \alpha \frac{(\Delta m_{ij}^2)^2 L}{E^2} \right)$ &
$\mathrm{GeV}^{-1} $ & 1 & 2 & 2  \\
 Ohlsson \cite{Ohlsson:2000mj} & $\exp \left( - \alpha \frac{(\Delta m_{ij}^2)^2 L^2}{E^2} \right)$ & dimensionless & 2 & 2 & 2  \\
\hline
\end{tabular}
\mycaption{\label{tab:qdecoherence} Different signatures that might
arise from quantum decoherence and the references in which they are
motivated.}
\end{center}
\end{table}

\subsubsection*{Other signatures}

In principle, what we have presented above is just a collection of
interesting signatures that could be responsible for damping of
neutrino oscillations. However, there are also other possibilities,
which we have decided not to investigate further in this study. These
signatures include, for example, heavy isosinglet neutrinos
\cite{Schechter:1980gr,Schechter:1980gk} and neutrino
oscillations in different extra dimension scenarios
\cite{Dvali:1999cn,Mohapatra:1999zd,Barbieri:2000mg,Mohapatra:2000wn,Morgan:2004vv,Hallgren:2004mw}.

\subsubsection*{Combined signatures}

In most cases, if there is a damping effect, then it would be natural
(and easy) to assume that one type of effect is giving a clearly
dominating contribution. However, if an experiment is carried out with
some specific setup, then contributions from different scenarios might
be of the same order. In such a case, the form of \equ{dfactor} is
spoiled. For example, in the case of neutrino decay combined with
neutrino absorption, the matrices $\Gamma$ are just added which
results in the damping signatures
\begin{equation}
D_{ij} = \exp\left[
-\left(\frac{\alpha_{ij}^{\rm decay}}{E} +
\alpha_{ij}^{\rm abs}E\right)L
\right].
\end{equation}
In general, just multiplying the damping factors (which is the result
of the above treatment) might not give the correct damping and
different combined cases might behave in other ways. However, since
there are different energy dependencies in the different damping
signatures, there will only be a limited energy range where a combined
treatment is necessary. In this study, we do not consider combined
signatures.

\section{Damped neutrino oscillation probabilities}
\label{sec:dampedprob}

In this section, we investigate the effects of damping on
specific neutrino oscillation probabilities interesting for
future reactor and long-baseline experiments, where we restrict
the analytical discussion to the vacuum case.

\subsection{The damped two-flavor neutrino scenario}
\label{sec:dampedtwoflavor}

In a simple two-flavor scenario, the damped neutrino oscillation
probabilities take particularly simple forms (just as in the non-damped
case). {}From the two-flavor equivalent of \equ{damping}, we obtain
\begin{eqnarray}
\label{equ:2flavs}
P_{\alpha\alpha} &=&
D_{11} c^4 + D_{22} s^4 + \frac{1}{2} D_{21} \sin^2(2\theta) \cos(2\Delta), \\
P_{\beta\beta} &=&
D_{11} s^4 + D_{22} c^4 + \frac{1}{2} D_{21} \sin^2(2\theta) \cos(2\Delta)
\end{eqnarray}
for the neutrino survival probabilities and
\begin{equation}
\label{equ:2flavo}
P_{\alpha\beta} = P_{\beta\alpha} =
\frac{1}{4} \sin^2(2\theta)[D_{11}+D_{22} - 2D_{21}\cos(2\Delta)]
\end{equation}
for the neutrino transition probability, where $\nu_\alpha$ is the
linear combination $\nu_\alpha = c \nu_1 + s \nu_2$, $\nu_\beta$ is
the linear combination that is orthogonal to $\nu_\alpha$, $\Delta
\equiv \Delta_{21}$, $s \equiv \sin(\theta)$, $c \equiv \cos(\theta)$, and
$\theta$ is the mixing angle between the two neutrino flavors.

Let us first discuss the case $\xi > 0$ or all $\alpha_{ii} = 0$,
which means that all $D_{ii}$ are equal to unity. We refer to this
case as ``decoherence-like'' (probability conserving) damping.  The
two-flavor formulas then become
\begin{equation}
\label{equ:2flavd}
P_{\alpha\beta} =
\delta_{\alpha\beta} + \frac{1}{2}(1-2\delta_{\alpha\beta})\sin^2(2\theta)
[1-D\cos(2\Delta)],
\end{equation}
where $D \equiv D_{21}$. Below, we will show that expressions
reminding of these two-flavor formulas will be quite common in the
three-flavor counterparts. In the limit $D \rightarrow 0$ (maximal
damping), the oscillations are averaged out, \ie,
$$
P_{\alpha\beta} \rightarrow
\delta_{\alpha\beta} [1 - \sin^2 (2 \theta)] + \frac{1}{2} \sin^2 (2 \theta),
$$
where the factor $1/2$ is typical for an averaged $\sin^2 (x)$
term. It is also of interest to note, from the form of \equ{2flavd},
that the neutrino transition probabilities can either be smaller or
larger than the undamped probabilities depending on the sign of
$\cos(2\Delta)$. For instance, the neutrino survival probability
\begin{equation}
P_{\alpha\alpha} =
  1 - \frac{1}{2}\sin^2 (2 \theta) [1 -  D \cos(2\Delta)].
  \label{equ:faketheta13}
\end{equation}
is smaller than the corresponding undamped probability if
$\cos(2\Delta)$ is positive and vice versa. Close to the oscillation
maximum $\Delta \sim \pi/2$, the factor $\cos(2\Delta)$ will be
negative, \ie, the damped neutrino survival probability will be larger
than the undamped probability, since the oscillations will be
partially averaged out. This behavior changes as a function of the
neutrino energy at points where $\cos(2\Delta)$ changes sign, \ie, at
$2\Delta = (n+1) \pi/2$, $n = 0,1,\hdots$. As a rule of thumb, the
damping will lead to larger probabilities close to the oscillation
maximum $E_{\mathrm{max}} = \Delta m^2 L/(2 \pi)$ and to smaller
probabilities for $E<2 E_{\mathrm{max}}/3$ and $E>2 E_{\mathrm{max}}$.
This result will be valid for any survival probability discussed in
this study.

{}From the form of \equ{2flavd}, it is apparent that if only a small
range of $\Delta$'s is studied, then a damping factor may mimic an
oscillation signal. The worst such case would be if the damping
signature had $\gamma = 2$. This would mean that if one makes a series
expansion of $\cos(2\Delta)$ and the exponential of the damping
factor, then the energy dependence will be the same to lowest order in
the expansion parameters, \ie, we will have
\begin{equation}
D\cos(2\Delta) = \left[1-\alpha|\Delta m^2|^\xi\frac{L^\beta}{E^2} + \ldots \right]
\left[1-\left(\frac{\Delta m^2L}{4E}\right)^2 + \ldots \right].
\end{equation}
This effect is also present in a general case with any number of
neutrino flavors.

Another interesting case is when $\alpha_{ij} = \alpha_i + \alpha_j$
and $\xi = 0$, which is expected for the neutrino decay and neutrino
absorption scenarios. This assumption results in the fact that the damping
factor $D_{ij}$ can be written as a product
\begin{equation}
\label{equ:pbviol}
D_{ij} = A_i A_j,
\end{equation}
where $A_i \equiv \exp(-\alpha_i L^\beta/E^\gamma)$ is only dependent on
the $i$th mass eigenstate. Then, the neutrino oscillation probabilities are
given by
\begin{eqnarray}
P_{\alpha\alpha}
&=&
A^2\left[(c^2+\kappa s^2)^2 - \kappa \sin^2(2\theta)
\sin^2(\Delta)\right], \\
P_{\beta\beta}
&=&
A^2\left[(\kappa c^2+s^2)^2 - \kappa \sin^2(2\theta)
\sin^2(\Delta)\right], \\
P_{\alpha\beta}
&=&
\frac{1}{4} A^2 \sin^2(2\theta)[1+\kappa^2 - 2\kappa \cos(2\Delta)],
\end{eqnarray}
where $A \equiv A_1$ and $\kappa \equiv A_2/A_1$. It is
important to note that, for example, the total probability
$P_{\alpha\alpha} + P_{\alpha\beta}$ is not conserved in this case, in
fact, we obtain
\begin{equation}
\label{equ:dprobtot}
P_{\alpha\alpha} + P_{\alpha\beta} =
A^2\left[c^4 + \kappa^2 s^4 +
\frac{1}{4}\sin^2(2\theta)(1+\kappa)^2\right] \leq 1,
\end{equation}
where the equality holds if and only if $A = \kappa = 1$ (because of
the form of the $A_i$'s, $A \leq 1$, $\kappa A \leq 1$, and that all
terms in \equ{dprobtot} are positive, the terms will attain their
maximum value when $A = \kappa A = 1$, in which case the entire
expression simplifies to one). Thus, we will introduce the term
``decay-like'' for effects giving rise to damping terms of the form
given in \equ{pbviol}.

In the case of a decay-like signature, there are two special cases
which are of particular interest. First, if both mass eigenstates are
affected in the same way, \ie, $\kappa = 1$, then the resulting
neutrino transition probabilities will reduce to the undamped standard
neutrino oscillation probabilities suppressed by a factor of
$A^2$. This means that all damped probabilities will be smaller than
their undamped counterparts. Second, if only one of the mass
eigenstates is affected, \ie, $A = 1$, then the difference in the
$\nu_\alpha$ survival probability compared to the undamped case will
be given by
\begin{equation}
\Delta P_{\alpha\alpha} \equiv P_{\alpha\alpha}^{\rm damped} -
P_{\alpha\alpha}^{\rm undamped} =
(\kappa-1) s^2 [(1+\kappa)s^2 + 2 c^2 \cos(2\Delta)].
\end{equation}
Thus, this survival probability will actually increase if
\begin{equation}
\label{equ:decincr}
- 2 \cos(2\Delta) > (1+\kappa)\tan^2(\theta).
\end{equation}
Note that for the first part of the neutrino propagation (for $L < \pi
E/\Delta m^2$), the term $\cos(2\Delta)$ is positive, and thus, the
inequality of \equ{decincr} cannot be satisfied in this region, since
the right-hand side is always positive. {}From the comparison with the
discussion after \equ{faketheta13}, this condition is equivalent to
$E>2 E_{\mathrm{max}}$. For example, for a neutrino factory, which can
be operated far away from the oscillation maximum, this implies that
the relevant part of the spectrum will be suppressed by this form of
damping. For the neutrino oscillation probability difference $\Delta
P_{\alpha\beta}$, we obtain
\begin{equation}
\Delta P_{\alpha\beta} = \frac 14 \sin^2(2\theta)(\kappa-1)
[1+\kappa - 2\cos(2\Delta)],
\end{equation}
that is, the damped $P_{\alpha\beta}$ is larger than the undamped
$P_{\alpha\beta}$ if
\begin{equation}
\label{equ:decincr2}
2\cos(2\Delta) > 1 + \kappa.
\end{equation}
Note that if $\tan(\theta) = 1$,
then \eqs~(\ref{equ:decincr}) and (\ref{equ:decincr2}) will have the same
form except for the sign of the left-hand side.

\begin{figure}[t]
\begin{center}
\includegraphics[width=10cm]{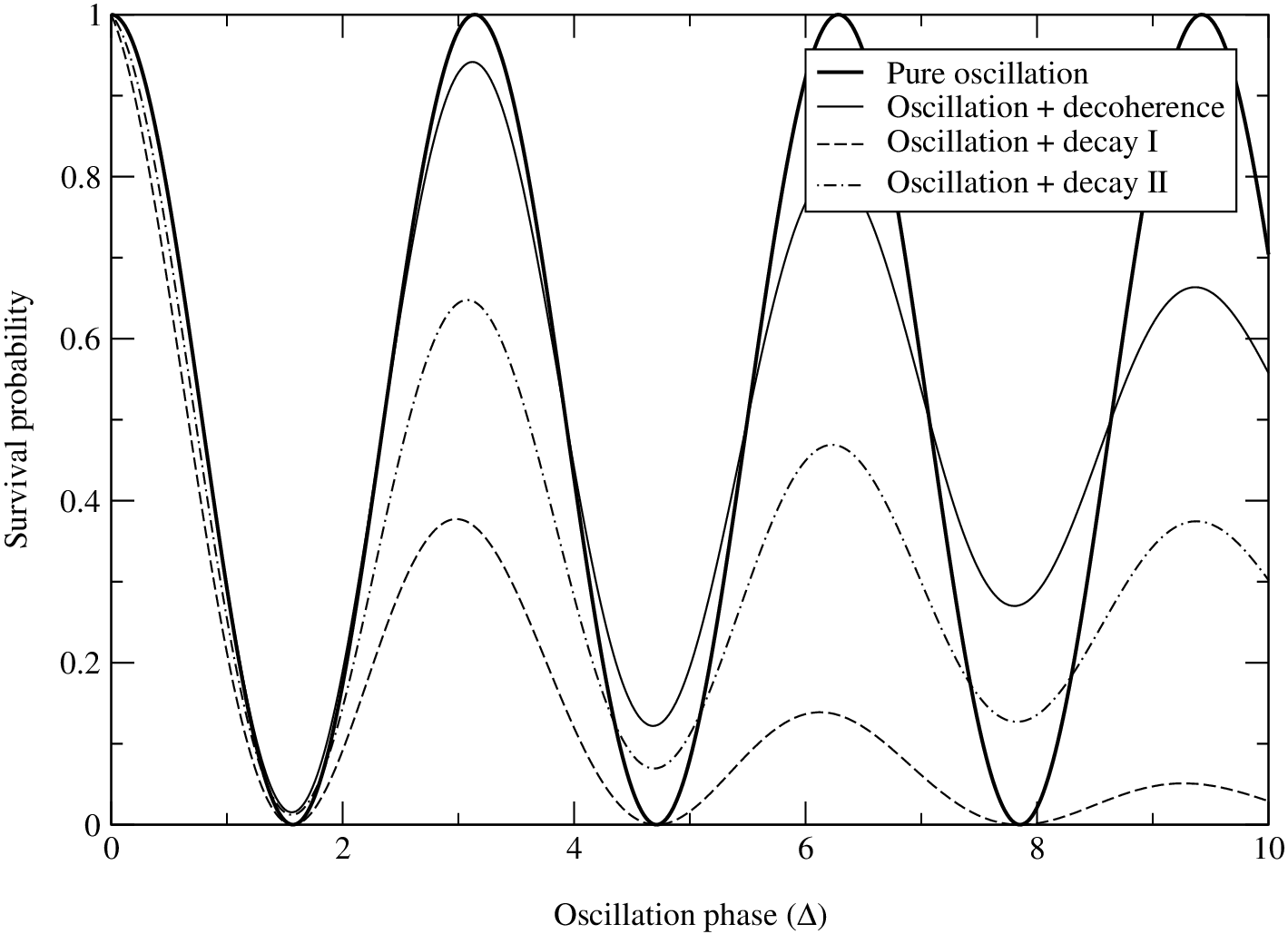}
\end{center}
\mycaption{\label{fig:illustration} The qualitative effect of
different damping signatures on the two-flavor neutrino survival
probability as a function of the oscillation phase $\Delta$. The
mixing used in this plot is maximal ($\theta = \pi/2$) and the
damping parameters have been highly exaggerated. The scenario
``Oscillation + decay I'' corresponds to decay of both mass
eigenstates with equal rates, whereas ``Oscillation + decay II''
corresponds to the second mass eigenstate decaying while the first
mass eigenstate is stable.}
\end{figure}

In \fig~\ref{fig:illustration}, the qualitative effects of neutrino
wave packet decoherence and neutrino decay on the neutrino survival
probability are shown.
{}From this figure, we clearly see how the wave packet decoherence
simply corresponds to a damping of the oscillating term and the decay
of all mass eigenstates corresponds to an overall damping of the
undamped neutrino survival probability. For the case of only one
decaying mass eigenstate, the probability converges towards the square of the
content of the stable mass eigenstate in the initial neutrino flavor
eigenstate.

\subsection{Three-flavor electron-muon neutrino transitions}

For a fixed neutrino oscillation channel, the damped neutrino
oscillation probability \equ{damping} can be written more explicitly
in terms of the mixing parameters and the mass squared
differences. Below, we will use the standard notation for the leptonic
mixing angles, \ie, $s_{ij} = \sin(\theta_{ij})$ and $c_{ij} =
\cos(\theta_{ij})$. Then, for example, the $\nu_e$ survival
probability $P_{ee}$ is given by
\begin{eqnarray}
P_{ee} &=&
c_{13}^4\left[
D_{11}c_{12}^4 + D_{22} s_{12}^4 +
\frac{1}{2} D_{21}\sin^2(2\theta_{12}) \cos(2\Delta_{21})\right] \nonumber \\
&&+ \frac{1}{2} \stheta
[D_{31}c_{12}^2 \cos(2\Delta_{31}) + D_{32} s_{12}^2 \cos(2\Delta_{32})]
+ D_{33} s_{13}^4,
\label{equ:Pee}
\end{eqnarray}
which is dependent on all neutrino oscillation parameters except for
$\theta_{23}$ and $\delta_{CP}$, while the probability $P_{e\mu}$ of
oscillations into $\nu_\mu$ is given by
\begin{eqnarray}
P_{e\mu} &=& \frac{1}{4} \sin^2(2\theta_{12})
c_{23}^2[(D_{11}+D_{22})-2D_{21}\cos(2\Delta_{21})] \nonumber \\ && +
\frac{1}{2}\sin(2\theta_{12})\sin(2\theta_{23})
\{c_\delta[D_{11}c_{12}^2-D_{22}s_{12}^2  - D_{21}
\cos(2\theta_{12})\cos(2\Delta_{21})] \nonumber \\ && -D_{21}s_\delta
\sin(2\Delta_{21}) +D_{32}\cos(2\Delta_{32} - \delta_{CP}) - D_{31}
\cos(2\Delta_{31} - \delta_{CP})\} \, s_{13}  \nonumber \\
&&
+s_{23}^2[D_{11} c_{12}^4 + D_{22} s_{12}^4
+ D_{33} - 2D_{31} s_{12}^2 \cos(2\Delta_{31}) -
2 D_{32} c_{12}^2 \cos(2\Delta_{32})] \, s_{13}^2  \nonumber \\
&& + \frac{1}{4} \sin^2(2\theta_{12}) [2 D_{21} \cos(2\Delta_{21})-
c_{23}^2(D_{11}+D_{22})] \, s_{13}^2  + \mathcal O(s_{13}^3),
\label{equ:Pemu}
\end{eqnarray}
where $s_\delta \equiv \sin(\delta_{CP})$ and $c_\delta \equiv
\cos(\delta_{CP})$. Furthermore, the $\nu_\mu$ survival probability
can be computed to be of the form
\begin{eqnarray}
P_{\mu\mu} &=&
\frac{1}{2} \sin^2(2\theta_{23})[D_{32}c_{12}^2 \cos(2\Delta_{32})
+D_{31}s_{12}^2 \cos(2\Delta_{31})] \nonumber \\
&&+c_{23}^4 \left[D_{11}s_{12}^4 + D_{22}c_{12}^4 + \frac{1}{2} D_{21}
\sin^2(2\theta_{12}) \cos(2\Delta_{21})\right] + D_{33}s_{23}^4
 \nonumber \\
&&+ c_\delta \sin(2\theta_{12})\sin(2\theta_{23})\left\{
c_{23}^2\left[
D_{11}s_{12}^2-D_{22}c_{12}^2+
{D_{21}}\cos(2\theta_{12})\cos(2\Delta_{21})
\right]\right. \nonumber \\
&&
\left.
+ s_{23}^2 [D_{31}\cos(2\Delta_{31}) - D_{32}\cos(2\Delta_{32})]
\right\} \, s_{13}
+\mathcal O(s_{13}^2).
\label{equ:Pmumu}
\end{eqnarray}
Note that the probabilities $P_{e\mu}$ and $P_{\mu\mu}$ are series
expansions in $s_{13}$, whereas the probability $P_{ee}$ is valid to
all orders in $s_{13}$. The reason to use these expressions rather
than the exact expressions is that, unless some further assumptions
are made, the formulas for $P_{e\mu}$ and $P_{\mu\mu}$ are quite
cumbersome.

The probability $P_{\mu e}$ can be obtained by making the
transformation $\delta_{CP} \rightarrow -\delta_{CP}$ in the
probability $P_{e\mu}$, \ie, $P_{\mu e} = P_{e\mu}(\delta_{CP}
\rightarrow -\delta_{CP})$. Furthermore, in vacuum, the anti-neutrino
oscillation probabilities can be obtain from the neutrino oscillation
probabilities through the same transformation as above. Note that this
is not true for neutrinos propagating in matter.

\subsection{Probabilities for decoherence-like effects in experiments}

For a decoherence-like damping effect, $D_{ii} = 1$ for all $i$ and
the relations
\begin{equation}
\label{equ:probconservation}
\sum_{\alpha = e,\mu,\tau} P_{\alpha\beta} = 1 \qquad {\rm and}
\qquad
\sum_{\beta = e,\mu,\tau} P_{\alpha\beta} = 1
\end{equation}
are still valid despite the presence of damping factors (\ie, no
neutrinos are lost due to effects such as invisible decay, absorption, \etc).
Note that, in the case of a
decoherence-like damping effect, all neutrino oscillation probabilities
can be constructed from $P_{ee}$, $P_{e\mu}$, and $P_{\mu\mu}$ due to
the conservation of total probability given in \equ{probconservation}.

It is interesting to observe what effect a decoherence-like damping
could have on the neutrino oscillation probabilities for different
experiments. Therefore, we will now study different kinds of neutrino
oscillation experiments and make different approximations depending on
the type of experiment to investigate what the main damping effects are.

\subsubsection*{Short-baseline reactor experiments}
\label{sec:sblreactor}

Short-baseline experiments, such as CHOOZ
\cite{Apollonio:1999ae,Apollonio:2002gd} and Double-CHOOZ
\cite{Ardellier:2004ui}, are operated at the atmospheric
oscillation maximum $\Delta_{31} \simeq \Delta_{32} = \mathcal{O}(1)$
in order to be sensitive to $\stheta$. The most interesting quantity
is the $\bar{\nu}_e$ survival probability $P_{\bar{e}\bar{e}}$. For
these experiments, it turns out (see \Sec~\ref{sec:appl1}) that it is
important to keep all damping factors. As a result, the $\bar \nu_e$
survival probability is given by
\begin{eqnarray}
P_{\bar{e}\bar{e}} &=&
c_{13}^4
\left\{1-\frac{1}{2}\sin^2(2\theta_{12})[1-D_{21}\cos(2\Delta_{21})]\right\}
\nonumber \\
&&
\label{equ:sblPee}
+ \frac{1}{2}\sin^2(2\theta_{13})
[D_{31}c_{12}^2 \cos(2\Delta_{31}) + D_{32} s_{12}^2 \cos(2\Delta_{32})]
+s_{13}^4.
\end{eqnarray}
The most apparent feature of this equation is the term within the
curly brackets, which has the form of the survival probability for a
two-flavor neutrino damping scenario with $\theta = \theta_{12}$ and
$\Delta = \Delta_{21}$. Therefore, even in the limit $\theta_{13}
\rightarrow 0$ [close to the $\stheta$ sensitivity limit], the damping
factor $D_{21}$ might be constrained by the contribution of the solar
oscillation at low energies.  Furthermore, in the limit $\Delta_{21}
\rightarrow 0$ (or large $\theta_{13}$), $D_{21}$ is close to unity
[\cf, \equ{coherence2}] and $D_{31} \simeq D_{32}$ (this could be
expected if $\Delta_{21}/\Delta_{31} \rightarrow 0$), then this
expression will exactly mimic the two-flavor neutrino damping scenario
with $\theta = \theta_{13}$ and $\Delta = \Delta_{31} = \Delta_{32}$.
Thus, depending on which small number (the ratio of the mass
squared differences or $s_{13}$) is the largest, two different
two-flavor neutrino scenarios are obtained as expected from the
non-damped case. If $\theta_{13}$ is relatively large (compared to the
ratio of the mass squared differences), then the latter two-flavor
case will apply. It is then interesting to note that the damping
factor $D_{31}$, the neutrino source energy spectrum, and the
cross-sections all have some energy dependence, which means that they
can ``emphasize'' certain regions in the energy spectrum which are
most sensitive to damping effects. If we assume that the total impact
is strongest close to the oscillation maximum, then the damping effect
will be misinterpreted as a smaller value of $\stheta$ [\cf, \equ{faketheta13},
which will in both cases be closer to unity]. Therefore, as we will
demonstrate, any such damping can fake a value of $\stheta$ which is
smaller than the one that is provided by Nature.

Note that, for the case of wave packet decoherence, $D_{21}$,
$D_{32}$, and $D_{31}$ are not independent [\cf, \equ{coherence2}],
which means that any of the terms in \equ{sblPee} could lead to
information on the parameter $\sigma_E$.

\subsubsection*{Long-baseline reactor experiments}

For long-baseline reactor experiments operated at the solar
oscillation maximum $\Delta_{21} = \mathcal{O}(1)$, such as the
KamLAND experiment \cite{Eguchi:2002dm,Araki:2004mb}, the damping
factors $D_{31}$ and $D_{32}$ of a decoherence-like scenario with $\xi
> 0$ are small, since the large mass squared
difference makes the argument of the exponential functions in
\equ{dfactor} large and negative. In addition, these two damping
factors are attached to neutrino oscillations associated with the
large phases $\Delta_{31}$ and $\Delta_{32}$ [see
\eqs~(\ref{equ:Pee})-(\ref{equ:Pmumu})], which effectively average
out. As a result of these two effects, the oscillating terms involving
the third mass eigenstate can be safely set to zero. After some
simplifications, the $\bar{\nu}_e$ survival probability
$P_{\bar{e}\bar{e}}$ is found to be
\begin{equation}
P_{\bar{e}\bar{e}}
=
c_{13}^4
\left\{1-\frac{1}{2}\sin^2(2\theta_{12})[1-D_{21}\cos(2\Delta_{21})]\right\}
+s_{13}^4. \label{equ:lblPee}
\end{equation}
This expression is clearly of the familiar form $P_{\bar{e}\bar{e}} =
c_{13}^4 P_{\bar{e}\bar{e}}^{\rm 2f} + s_{13}^4$, where
$P_{\bar{e}\bar{e}}^{\rm 2f}$ is the damped two-flavor $\bar\nu_e$
survival probability with $\theta = \theta_{12}$ and $\Delta =
\Delta_{21}$, which is also obtained in the non-damped case when
averaging over the fast oscillations [\cf~\equ{sblPee}]. For the case
of wave packet decoherence, we know from \equ{coherence2} that the
parameter $\sigma_E$ could be constrained by either of these two
equations. Since this parameter is experiment dependent, one could
argue that one should obtain some limits from the KamLAND experiment,
because the reactor experiments are very similar in source and
detector (see, \eg, \Ref~\cite{Schwetz:2003se}). However, it should be noted
that KamLAND has a rather weak precision on the corresponding
$\theta_{12}$ measurement because of normalization
uncertainties. Since a decoherence contribution would appear at low
energies, the data set in \Ref~\cite{Araki:2004mb} does not seem to be
very restrictive for the parameter $\sigma_E$.

\subsubsection*{Beam experiments}

For beam experiments, such as superbeams, beta-beams or neutrino
factories, one may assume $\Delta_{21} \simeq 0$ as a first
approximation if one wants to be sensitive to $\stheta$, since, at the
energies and baseline lengths involved, the low-frequency neutrino
oscillations do not have enough time to evolve. In the case of $\xi
> 0$, this also implies that $D_{12} = 1$ and $D \equiv D_{32} =
D_{31}$ to a good approximation. {}From these assumptions, it follows that
\begin{eqnarray}
P_{e\mu} &=& 2 s_{23}^2 [1 - D \cos(2\Delta)] \, s_{13}^2 + \mathcal O(s_{13}^3), \\
P_{\mu\mu} &=& 1 - \frac{1}{2}\sin^2(2\theta_{23}) [1 - D \cos(2\Delta)] +
\mathcal O(s_{13}^2),
\label{equ:pmumudamped}
\end{eqnarray}
where $\Delta \equiv \Delta_{32} = \Delta_{31}$. Note that the
probability $P_{e\mu}$ is correct up to $\mathcal O(s_{13}^3)$ [as
compared with \equ{Pemu}, which is only valid up to $\mathcal
O(s_{13}^2)$], this is one of the cases where the assumptions made
simplifies the $s_{13}^2$ term in this probability. Both of the above
equations show obvious similarities with the cases of damped
two-flavor neutrino oscillations. For $P_{e\mu}$ we have an
approximate two-flavor neutrino scenario with $s^2c^2 = s_{23}^2
s_{13}^2$ and $P_{\mu\mu}$ is a pure two-flavor neutrino formula with
$\theta = \theta_{23}$ up to the corrections of order $s_{13}^2$.
Since the disappearance channel $P_{\mu\mu}$ at a beam experiment is
supposed to have extremely good statistics, $D$ will be strongly
constrained by this channel.  Note that the damping in $P_{\mu\mu}$
qualitatively behaves as the one in \equ{faketheta13}, \ie, the damped
probability might be larger or smaller than the undamped probability
depending on the position relative to the oscillation maximum
$E_{\mathrm{max}}$.

\subsection{Probabilities for decay-like effects in experiments}

If $\xi = 0$ and $\alpha_{ii} \neq 0$, then $D_{ii} \neq 1$ and
\equ{probconservation} will not hold. We define any effect of this
kind to be ``probability violating''. As mentioned in the two-flavor
neutrino discussion, a very interesting special case of the
probability violating effects is the case of a decay-like effect. The
neutrino oscillation probabilities for decay-like effects
corresponding to the ones given for decoherence-like effects are
listed below.

\subsubsection*{Short-baseline reactor experiments}

For the short-baseline reactor experiments, we obtain the
$\bar\nu_e$ survival probability as
\begin{eqnarray}
P_{\bar e\bar e} &=&
c_{13}^4 \left\{
(A_1 c_{12}^2 + A_2 s_{12}^2)^2 -
A_1 A_2 \sin^2(2\theta_{12})\sin^2(\Delta_{21})
\right\} \nonumber \\
&&
+A_3 s_{13}^2\{A_3 s_{13}^2 +
2 c_{13}^2[A_1 c_{12}^2 \cos(2\Delta_{31})+A_2s_{12}^2\cos(2\Delta_{32})]\}.
\end{eqnarray}
Again, as in the case of decoherence-like damping, the
expression within the curly brackets is of a two-flavor form with
$\theta = \theta_{12}$ and $\Delta = \Delta_{12}$. In the limit when
$\stheta$ is large and we ignore the solar oscillations, we obtain the
two-flavor neutrino scenario
\begin{equation}
P_{\bar e\bar e} = A^2\left\{
(c_{13}^2+\kappa s_{13}^2)^2 - \kappa \stheta \sin^2(2\Delta)
\right\}
\end{equation}
only if we assume that $A_1 = A_2 = A$, where $\Delta = \Delta_{31} =
\Delta_{32}$ and $\kappa = A_3/A$.

\subsubsection*{Long-baseline reactor experiments}

Assuming that the fast neutrino oscillations average out, the
$\bar\nu_e$ survival probability is given by
\begin{equation}
P_{\bar e\bar e} = c_{13}^4 P_{\bar e\bar e}^{\rm 2f} + A_3^2 s_{13}^4,
\end{equation}
where $P_{\bar e\bar e}^{\rm 2f}$ is the two-flavor decay-like
$\bar\nu_e$ survival probability with $\theta = \theta_{12}$ and
$\Delta = \Delta_{21}$ [\cf, \eq~(\ref{equ:lblPee})]. In this
expression, the $s_{13}^4$ term is also damped, which does not apply
in a decoherence-like scenario.

\subsubsection*{Beam experiments}

When the assumptions $\Delta_{21} \simeq 0$ and $A = A_1 = A_2$ (which
could be expected in a decay scenario where $m_1 = m_2$) are made, the
neutrino oscillation probabilities that are relevant for beam
experiments become
\begin{eqnarray}
P_{e\mu}
&=&
A^2 s_{23}^2 [1 + \kappa^2 - 2\kappa \cos(2\Delta)] \, s_{13}^2
+ \mathcal O(s_{13}^3), \\
P_{\mu\mu}
&=&
A^2\left[(c_{23}^2 + \kappa s_{23}^2)^2 -
\kappa\sin^2(2\theta_{23}) \sin^2(\Delta)\right] + \mathcal O(s_{13}^2),
\end{eqnarray}
where $\kappa \equiv A_3/A$ and $\Delta \equiv \Delta_{32} =
\Delta_{31}$. These probabilities mimic decay-like two-flavor
probabilities just as the corresponding decoherence-like effects
mimic decoherence-like two-flavor probabilities to leading order
in $s_{13}$.

\section{Application I: Faking a small $\boldsymbol{\stheta}$ at reactor experiments by decoherence-like effects}
\label{sec:appl1}

In this section, we demonstrate the possible effects of damping at a
simple example using a full numerical simulation. Let us only consider
the case of intrinsic wave packet decoherence, which is very
interesting from the point of view that it is a ``standard'' effect in
any realistic neutrino oscillation treatment. However, similar effects
could occur from related signatures, such as quantum decoherence. As
experiments, one could, in principle, consider all classes of
experiments in order to investigate decoherence signals.  New reactor
experiments with near and far
detectors~\cite{Minakata:2002jv,Huber:2003pm} are candidates for
``clean'' measurements of $\stheta$, \ie, they are specifically
designed to search for a $\stheta$ signal.  As we have discussed in
\Sec~\ref{sec:sblreactor}, an interesting decoherence-like effect at
such an experiment would be a derived value of $\stheta$ which is
smaller than the value provided by Nature. In this case, the CHOOZ
bound might actually be too strong and the interpretation of new
reactor experiments might be wrong.

\begin{figure}[t]
\begin{center}
\includegraphics[width=\textwidth]{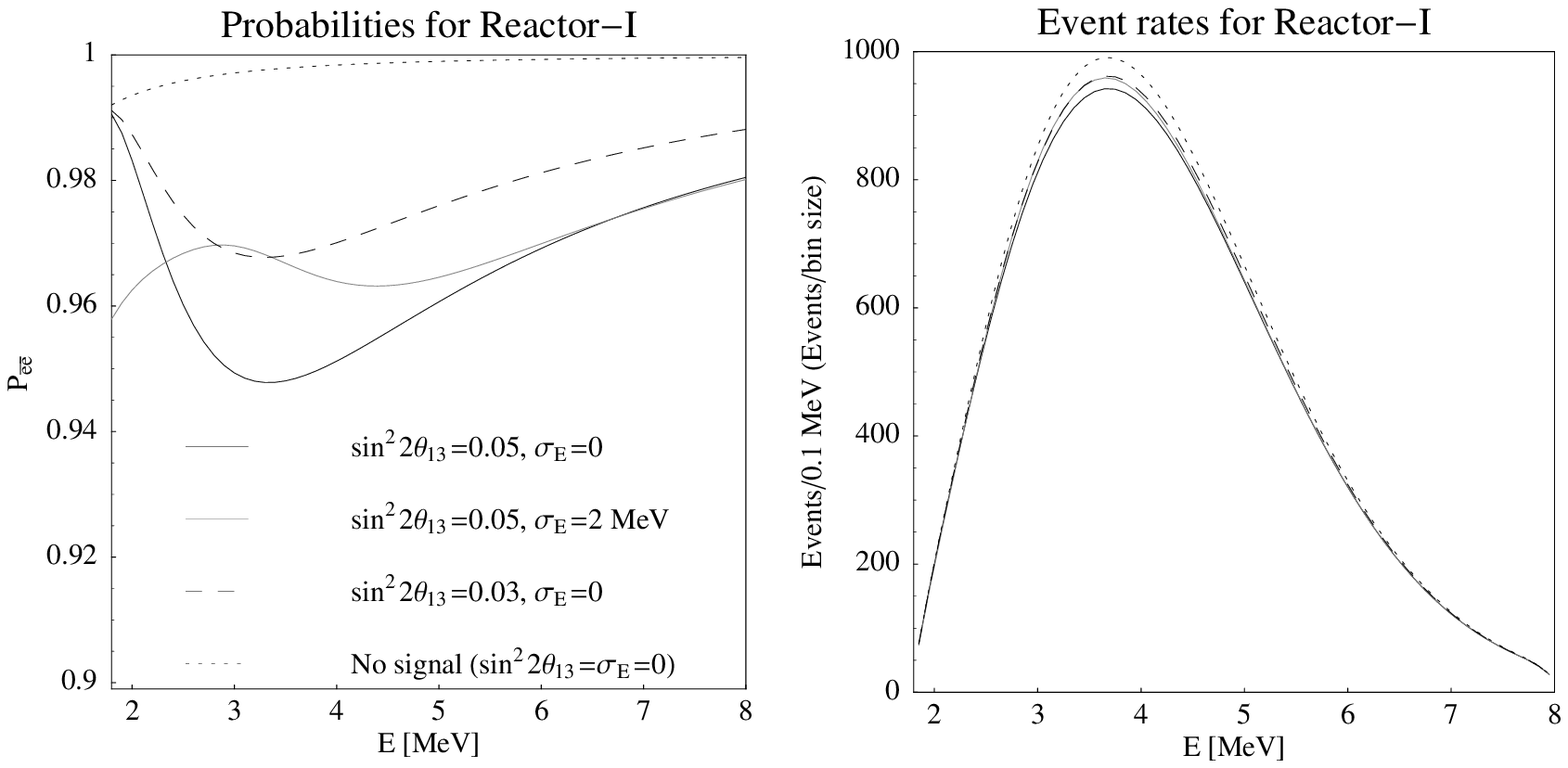}
\end{center}
\mycaption{\label{fig:reactorprobs} The neutrino oscillation
probability $P_{\bar{e} \bar{e}}$ (left) and event rates (right) for
the experiment \ReactorI\ from \Ref~\cite{Huber:2003pm} in the
analysis range. For the simulated parameter values, we use $\ldm = 2.5
\cdot 10^{-3} \, \mathrm{eV}^2$, $\sdm = 8.2 \cdot 10^{-5} \,
\mathrm{eV}^2$, $\sin^2 2 \theta_{12}=0.83$, $\sin^2 2 \theta_{23}=1$,
$\delta_{CP}=0$~\cite{Fogli:2003th,Bahcall:2004ut,Bandyopadhyay:2004da,Maltoni:2004ei}
and the values for $\stheta$ and $\sigma_E$ as given in the left
plot.}
\end{figure}

If we assume that there is an intrinsic loss of coherence, then the
reactor $\bar \nu_e$ survival probability $P_{\bar{e} \bar{e}}$ will
be given by \equ{sblPee}. In order to illustrate the decoherence
effect, we show in \figu{reactorprobs} $P_{\bar e \bar e}$ and the
corresponding event rates for the experiment \ReactorI\ from
\Ref~\cite{Huber:2003pm} (full analysis range shown).
The different curves correspond to the non-oscillatory case as well as
different combinations of $\stheta$ and $\sigma_E$. As one can
observe, the two cases $\stheta$ large and decoherence [$\stheta=0.05$
and $\sigma_E=2 \, \mathrm{MeV}$] and $\stheta$ small and no
decoherence [$\stheta=0.03$ and $\sigma_E=0$] correspond, especially
in the event rate plot, very well to each other [as compared to the
other two cases of no oscillations and large $\stheta$ only]. This
means that the decoherence effect can mimic a smaller value of
$\stheta$ than what is provided by Nature. Note that in the
probability plot, there is a significant contribution from loss of
coherence in the solar terms for low energies. As we will see later,
this contribution can limit the decoherence effects even for no
$\stheta$ signal. In addition, the damped neutrino oscillation
probability is larger than the undamped one in the range discussed
after \equ{faketheta13}, where the oscillation maximum is here at
about $E_{\mathrm{max}} \simeq 3.4 \, \mathrm{MeV}$.

\begin{figure}[t]
\begin{center}
\includegraphics[width=\textwidth]{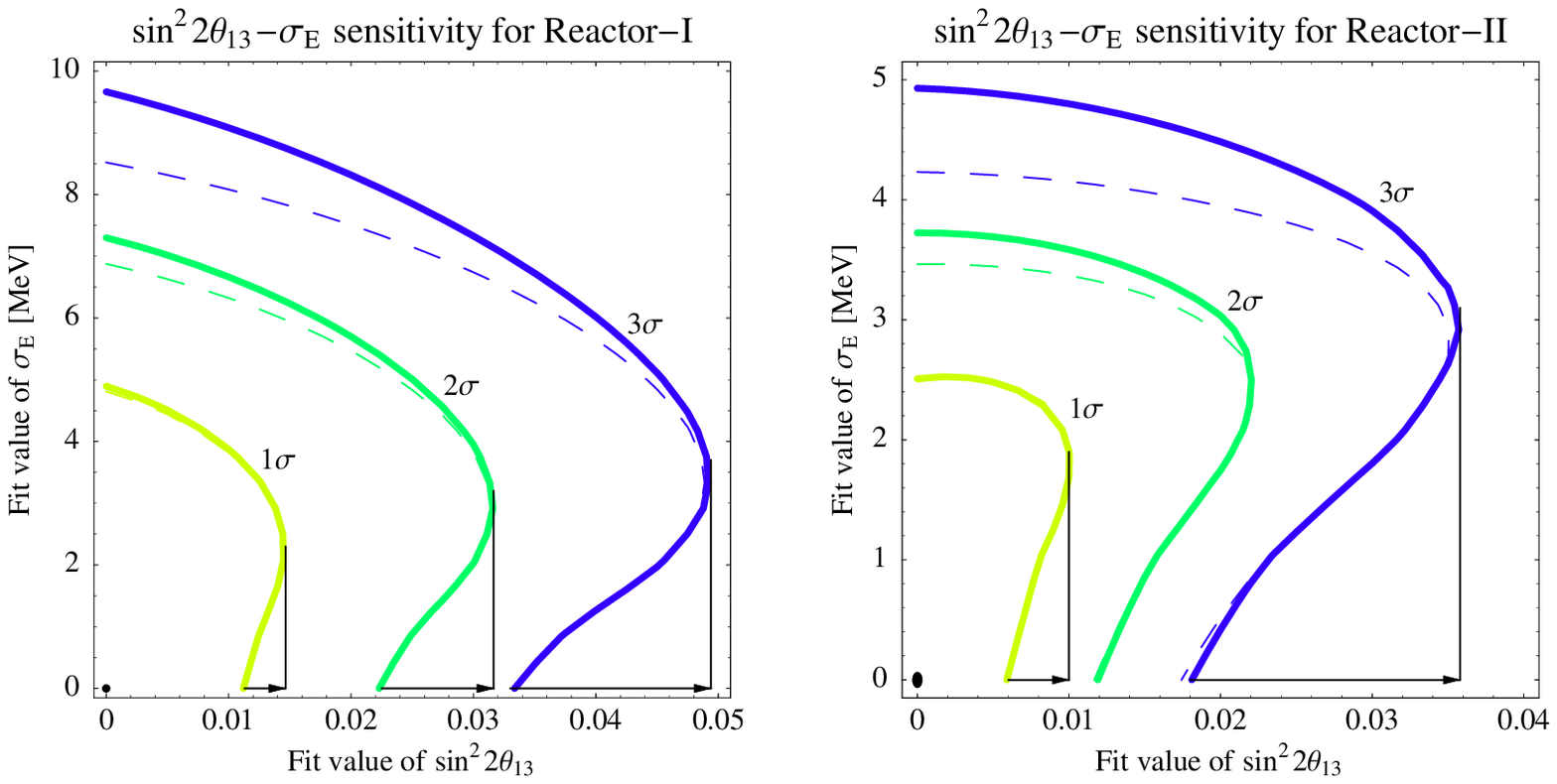}
\end{center}
\mycaption{\label{fig:reactorcorr} Simultaneous sensitivity to
$\stheta$ and $\sigma_E$ for the experiments \ReactorI\ (left) and
\ReactorII\ (right) from \Ref~\cite{Huber:2003pm} (curves shown for 1
d.o.f). For the simulated parameter values, we use $\stheta=0$,
$\sigma_E=0$, and the other values as in \figu{reactorprobs}.  For the
thick solid curves, the unshown fit parameter values are marginalized
over, where post-KamLAND external precisions of 10~\% on each $\sdm$
and $\theta_{12}$~\cite{Gonzalez-Garcia:2001zy,Barger:2000hy} are
imposed along with an external error of 10~\% for $\ldm$ obtained from
the superbeams is assumed. For the thin dashed curves, the unshown fit
parameter values are fixed (no correlations).  For the numerical
analysis, an extended version of the GLoBES
software~\cite{Huber:2004ka} is used. The arrows indicate the shift of
the $\stheta$ sensitivity limit if one assumes $\sigma_E$ as a free
parameter.}
\end{figure}

In order to illustrate the effect for a complete analysis, we show in
\figu{reactorcorr} the simultaneous sensitivity to $\stheta$ and
$\sigma_E$ for \ReactorI\ ($\mathcal{L} = 400 \, \mathrm{t \, GW \,
yr}$) and \ReactorII\ ($\mathcal{L} = 8 \, 000 \, \mathrm{t \, GW \,
yr}$) from \Ref~\cite{Huber:2003pm} (1 d.o.f.)  using an extended
version of the GLoBES software~\cite{Huber:2004ka}. In this figure,
$\sigma_E$ is assumed to be a free (fit) parameter that has to be
measured by the experiment.
Therefore, without additional knowledge, the
$\stheta$ sensitivity limit is obtained as a projection of the curves
onto the $\stheta$-axis. Since the $\stheta$ sensitivity limit for no
decoherence effects is the one for $\sigma_E=0$, the arrows indicate
the shift of this limit by the unknown $\sigma_E$.  This means, for
example, that the sensitivity limit becomes about 50~\% to 100~\%
worse than that for the actual $\sigma_E \equiv 0$, since the
decoherence mimics a smaller value of $\stheta$ than what is provided
by Nature. Similar results to the left plot are obtained for the
proposed Double-CHOOZ experiment~\cite{Ardellier:2004ui}. Note that
the correlation between $\stheta$ and $\sigma_E$ affects the $\stheta$
sensitivity (projection onto the horizontal axis), but not the
$\sigma_E$ sensitivity (projection onto the vertical axis). The latter
is correlated with the other neutrino oscillation parameters
(especially the solar parameters), as one can read off from the
difference between the solid and dashed curves. For the $\sigma_E$
sensitivity, one obtains $\sigma_E \lesssim 10 \, \mathrm{MeV}$
(\ReactorI ) and $\sigma_E \lesssim 5 \, \mathrm{MeV}$ (\ReactorII )
at the $3 \sigma$ confidence level.
As one can observe from the left plot of
\figu{reactorprobs}, there is some contribution of the solar
oscillation averaging to the decoherence effect at low energies.  In
fact, this is the reason why one can constrain $\sigma_E$ even for
$\stheta \equiv 0$, since in the decoherence effect, the atmospheric
oscillations are suppressed by the oscillation amplitude
$\stheta$. Obviously, this solar decoherence effect determines the
upper bound for $\sigma_E$, which means that the $\sigma_E$
sensitivity is limited by the knowledge on the solar oscillation
parameters [\cf, \equ{sblPee}].

As we have discussed in \Sec~\ref{sec:phenomenology}, $\sigma_E$ might
be an experiment dependent parameter related to the production and
detection processes. Instead of deriving bounds for this parameter
from reactor experiments, one can estimate from \figu{reactorcorr}
that one has to constrain $\sigma_E$ better than to about $\sigma_E
\lesssim 0.5 \, \mathrm{MeV}$ in order not to have a significant
deterioration of the $\stheta$ sensitivity limit.
In addition, in order to exclude an experiment dependent effect, it is highly
recommendable to measure the same quantity with different techniques
such as $\stheta$ with both reactor experiments and superbeams.

\section{Application II: Testing and disentangling damping signatures at neutrino factories}
\label{sec:appl2}

If we want to constrain the model parameters in \Tab~\ref{tab:models}
and to test the different models against each other, then we will need
to choose a high-precision instrument to test these tiny effects.
Therefore, we investigate the potential of a neutrino factory.
In particular, the muon neutrino disappearance channel $\nu_\mu
\rightarrow \nu_\mu$ at a neutrino factory has very good statistics
and the impact of neutrino oscillation parameter correlations other
than with $\ldm$ and $\theta_{23}$ is very small. Thus, we will mainly
focus on this disappearance channel, but include the appearance
information in the full analysis and demonstrate how the value of
$\stheta$ would influence the effects.  Since our exponential damping
model is not directly comparable to other approaches in the
literature, we put a major emphasis on the identification problem of a
non-standard contribution: If we actually observe something
unexpected, how well can we determine what sort of effect this
actually is? In the simplest case, this means that we test a signature
against the standard (no damping) scenario giving us limits for the
parameters. Since it is almost impossible to include the correlations
among all parameters, we choose to use $\alpha_{ij}=\alpha$
independent of $i$ and $j$ in this section in order to drastically
reduce the number of parameters. This means that we now have to deal
with eight correlated parameters (six neutrino oscillation parameters,
the matter density, and the parameter $\alpha$). We have motivated
this choice at the end of \Sec~\ref{sec:gendescription} and, for
individual cases, in \Sec~\ref{sec:examples}.

\begin{figure}[t]
\begin{center}
\includegraphics[width=\textwidth]{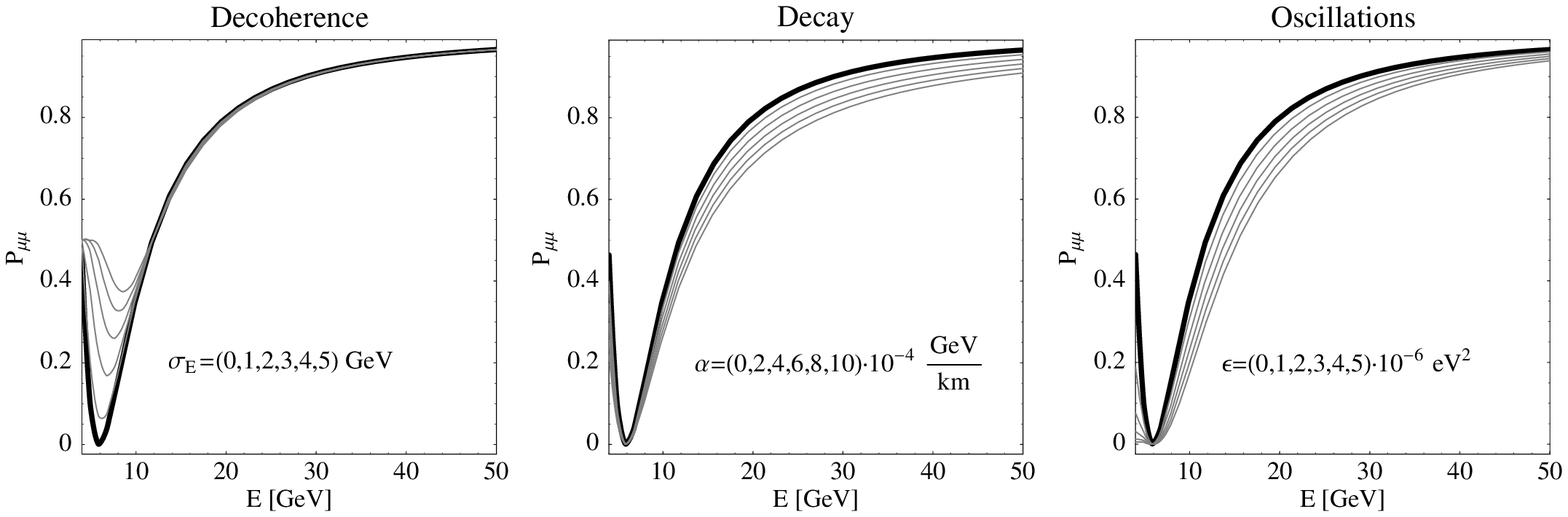}
\end{center}
\mycaption{\label{fig:allprobs} Contributions of the first three
different damping signatures from \Tab~\ref{tab:models} to the
disappearance probability $P_{\mu \mu}$ as function of the neutrino
energy. Here $L=3 \, 000 \, \mathrm{km}$ and the neutrino oscillation
parameters as in \figu{reactorprobs} with $\stheta=0$ are used. The
parameters for the non-standard effects are given in the plots, where
zero corresponds to the thick curves (oscillations only) and larger
values correspond to curves further off the zero curve. The energy
range corresponds to the analysis range of the $50 \, \mathrm{GeV}$
neutrino factory \NuFactII\ from \Ref~\cite{Huber:2002mx}.}
\end{figure}

Before we come to the results of a complete simulation, let us
illustrate the spectral behavior (energy dependence) of the neutrino
oscillation probability $P_{\mu \mu}$ in \figu{allprobs} for some
characteristic examples.  Earlier in \Sec~\ref{sec:dampedprob}
we have already discussed that there
are two general interesting cases: Either only the oscillatory terms
are damped or all terms are damped. In \figu{allprobs}, we can clearly
identify this difference between the decoherence-like and the other
two damping models (decay and oscillations). In all the shown cases
(for which $\gamma>0$), the relative importance of the damping
increases as the energy decreases.  However, since also the neutrino
oscillation probability drops with lower energies, the absolute size
of the effect is determined by the ratio of signature versus
probability effect for low energies. In addition, cross-section and
flux will disfavor low energies, which means that the low-energy
effects become even harder to identify.  This makes the wave packet
decoherence scenario most difficult to test, since the $E^{-4}$
dependence in the exponent strongly favors low energies. However, it
might be most easily distinguished from the decay and oscillation
damping scenarios because of its unique signature.  As we have
discussed after \equ{faketheta13} [which also holds for the similar
\equ{pmumudamped}], it is a characteristic feature of decoherence-like
signatures that they cross the undamped curve at $2E_{\mathrm{max}}/3$
and $2E_{\mathrm{max}}$, which here evaluate to $4 \, \mathrm{GeV}$
(outside of the analysis range) and $12 \, \mathrm{GeV}$. In
\figu{allprobs} (left panel), this effect is hardly observable because
of the $E^{-4}$ energy dependence, but the quantum decoherence
motivated case ``Quantum decoherence II'' from \Tab~\ref{tab:models}
clearly shows this behavior because of an $E^{-2}$ energy dependence.
As far as the other two signatures are concerned, the decay damping
has a linear energy dependence in the exponent as opposed to the
quadratic one for the oscillation damping scenario. Therefore, one has the
strongest high-energy effect for the decay damping scenario.

\begin{table}[t]
\begin{center}
{\footnotesize
\begin{tabular}{l|cccccc}
\hline
 & \multicolumn{6}{c}{Simulated damping signature} \\
\hline
 & Decoherence & Decay & Oscillations & Absorption & Q.~decoh.~I & Q.~decoh.~II \\
Fit signature & $\frac{\sigma_E}{\mathrm{GeV}}$ $\gtrsim$  & $\frac{\alpha}{10^{-5} \, \mathrm{\frac{GeV}{km}}}\gtrsim$ & $\frac{\epsilon}{10^{-7} \, \mathrm{eV}^4} \gtrsim$ &
$\frac{\alpha}{\frac{10^{-8}}{\mathrm{GeV \, km}}} \, \gtrsim$ & $\frac{\alpha}{\frac{10^{-10}}{\mathrm{GeV^2 \, km}}} \gtrsim$ & $\frac{\kappa}{\frac{10^{24}}{ \mathrm{eV}^2}} \gtrsim$
\\[3mm]
\hline
No damping & 1.7 (2.8) & 4.3 (7.2) & 5.1 (8.3) & 1.9 (3.1) & 4.1 (6.8) & 2.0 (3.6) \\
\hline
Decoherence & - & 4.3 (7.2) & 5.1 (8.3) & 1.9 (3.1) & 4.1 (6.8) & 2.0 (3.6) \\
Decay & 1.7 (2.8) & - & 6.3 (10) & 3.4 (5.7) & 6.0 (10) & 2.6 (5.1) \\
Oscillations & 1.7 (2.8) & 5.8 (9.8) & - & 1.9 (3.2) & 4.1 (6.9) & 13 (17) \\
Absorption & 1.7 (2.8) & 7.8 (13) & 5.2 (8.5) & - & 24 (40) & 2.1 (3.8) \\
Q.~decoh.~I & 1.7 (2.8) & 6.3 (11) & 5.1 (8.3) & 11 (19) & - & 2.1 (3.7) \\
Q.~decoh.~II & 1.7 (2.8) & 4.3 (7.2) & 5.1 (8.3) & 1.9 (3.1) & 4.1 (6.8) & - \\
\hline
All models & 1.7 (2.8) & 7.8 (13) & 6.3 (10) & 11 (19) & 24 (40) & 13 (17) \\
\hline
\end{tabular}
}
\end{center}
\mycaption{\label{tab:results} Parameter sensitivity limits for which
the simulated models (in columns) from \Tab~\ref{tab:models} could be
distinguished from the fit models (in rows) at the $3 \sigma$ ($5
\sigma$) confidence level (for the experiment simulation \NuFactII\
from \Ref~\cite{Huber:2002mx}). For example, decoherence could be
established against all models (including standard oscillations) for
the simulated $\sigma_E \gtrsim 1.7 \, \mathrm{GeV}$. For the simulated
neutrino oscillation parameter values, we use the same values as in
\figu{reactorprobs} and $\stheta=0$ as given in the column
captions. The fit parameter values (including the model parameter
$\alpha$) are marginalized over. The row ``no damping'' corresponds to
the standard neutrino oscillation scenario, \ie, it corresponds to the
upper bounds for the parameters assuming that there is only one
non-standard effect.  The row ``All models'' corresponds to the most
conservative case, \ie, it is an estimate for how well one can
establish the model against all of the other shown models.}
\end{table}

In order to test the different models against each other, we use a
modified version of the GLoBES software~\cite{Huber:2004ka} and the
neutrino factory setup \NuFactII\ from \Ref~\cite{Huber:2002mx}. This
neutrino factory uses a $50 \, \mathrm{kt}$ magnetized iron detector,
$4 \, \mathrm{year}$ of running time in each polarity, and $4 \,
\mathrm{MW}$ target power (corresponding to $5.3 \cdot 10^{20}$ useful
muon decays per year). For a fixed set of simulated parameter values
including the simulated damping parameter $\alpha$, we marginalize
over the fit neutrino oscillation parameters including the fit damping
parameter.  Due to the complexity of the parameter space, we assume
that the $\mathrm{sgn}(\ldm)$-degeneracy has been resolved by this
time.  We define the sensitivity limit to $\alpha$ as the threshold
above which the simulated damping model could be distinguished from
the fit damping model. Thus, if the damping mechanism is really there,
then the damping parameter $\alpha$ has to be above this threshold in
order to establish the model against the fit model with the given
experiment. In particular, we include the fit damping model ``no
damping'', which corresponds to the standard neutrino oscillation
case. For the simulation, we impose external precisions of 10~\% on
each $\theta_{12}$ and
$\sdm$~\cite{Gonzalez-Garcia:2001zy,Barger:2000hy}.  In addition, we
assume a constant matter density profile with 5~\% uncertainty, which
takes into account matter density uncertainties as well as matter
density profile effects~\cite{Geller:2001ix,Ohlsson:2003ip,Pana}.
However, we assume that the neutrino factory itself measures $\ldm$
and $\theta_{23}$ with its disappearance channel, \ie, we do not
impose an external precision on these parameters.

The resulting sensitivity limits of this analysis are shown in
\Tab~\ref{tab:results}, where the columns correspond to the simulated
models and the rows correspond to the fit models. These results are
computed for $\stheta=0$. It turns out that for a simulated value of
$\stheta$ close to the CHOOZ bound $\stheta \simeq 0.1$, the limits on
$\alpha$ would improve up to about 30~\% [depending on model and value
of $\stheta$] because of the additional contribution from the
appearance signal.\footnote{We do not show these results, since the
exact interpretation of the appearance signal is model dependent.  In
addition, matter effects are strong in this case and they depend on
the treatment of those in the context of the damping model.} Let us
first of all discuss the resulting sensitivities against the standard
neutrino oscillation scenario for some simple cases. For decoherence,
the obtained numbers indeed correspond very well to the energy
resolution of the detector, which is about 15~\% of the neutrino
energy, \ie, $1.5 \, \mathrm{GeV}$ for a neutrino energy of $E=10 \,
\mathrm{GeV}$, where the major effect takes place (\cf,
\figu{allprobs}, left plot).  Since the neutrino oscillation
probability changes sufficiently fast in this region, the measurement
is limited by the energy resolution of the detector.
In the wave packet approach, the bound against the ``no damping'' model
$\sigma_E \lesssim 1.67 \, \mathrm{GeV}$ translates into $\sigma_x
\gtrsim 6 \cdot 10^{-17} \, \mathrm{m}$. This rather small number
(sub-nucleon size) means that the bound is not very useful for wave
packet decoherence, since it is virtually impossible to create such
sharply peaked wave packets. However, there might be other energy
averaging effects that can be constrained.  For decay, we obtain a
limit, against the standard model, which is comparable to the current
neutrino lifetime limit for $m_3$.  Note that we have included all
correlations with the neutrino oscillation parameters in this
limit. However, the limit would be a factor of two weaker if we
considered only decay of $m_3$ instead of all mass eigenstates. Since
there are quite strong bounds on the $m_1$ and $m_2$ lifetimes from
supernova and solar neutrino observations, this factor of two
difference should be a very good approximation for the actual limit.
For the oscillation signature, the obtained limits are of the order of
magnitude $5 \cdot 10^{-7} \, \mathrm{eV}^2$, which corresponds to
$(\Delta m_{43}^2)^2$ times the active-sterile mixing in our estimate
for a possible mass scheme (\cf, \Sec~\ref{sec:oscsteriles}).
Considering the $\Delta m_{43}^2$ dependence, this is in fact not a
very strong bound. However, note that we have taken into account the
full parameter correlation, \ie, this effect could not come from
$\stheta$ or any other standard parameter.

\begin{figure}[t]
\begin{center}
\includegraphics[width=10cm]{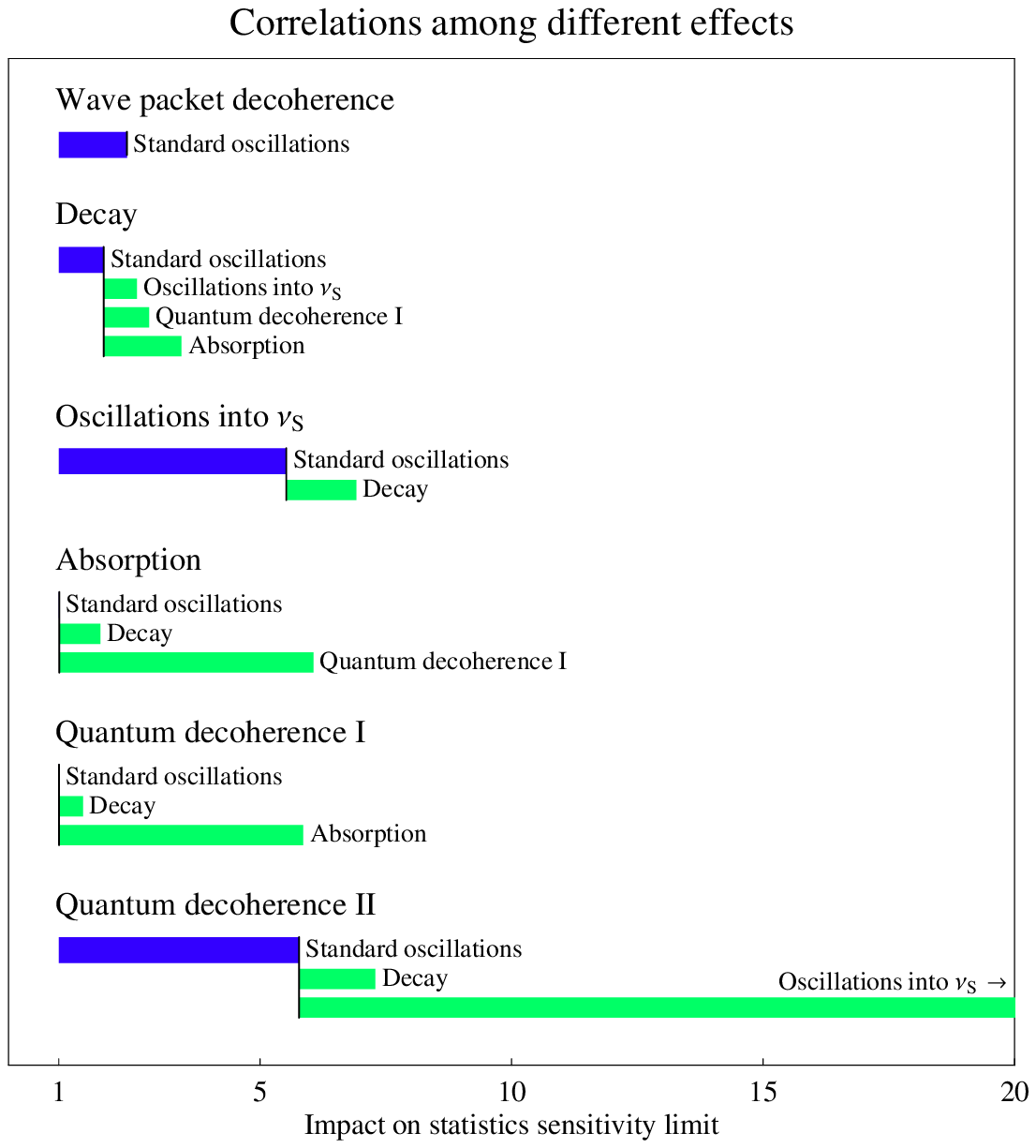}
\end{center}
\caption{\label{fig:corrplot} The impact of different correlations on
the statistics (and systematics) sensitivity limit of the model
dependent parameter $\alpha$ ($3 \sigma$), where the horizontal axis
represents multiples of the statistics (and systematics) sensitivity
limit. The group captions refer to the simulated models and the bar
labels to the fit models, where only the fit models are shown which
affect the sensitivity limit more than by 5~\%. The dark bars represent
the correlations with the neutrino oscillation parameters (fit parameter
$\alpha=0$ fixed) and the light bars indicate the additional change
if the model specific fit parameter $\alpha$ is marginalized over. The
lowest light bar extends to $37$.}
\end{figure}

In order to discuss the general identification problem among different
damping signatures, some information can be obtained from
\Tab~\ref{tab:results}. In addition, in \figu{corrplot} we show the
impact of the correlations with the standard neutrino oscillation
parameters (dark bars) as well as the additional correlation with the
fit model parameter $\alpha$ (light bars) on the $\alpha$ sensitivity
limit for the simulated models from \Tab~\ref{tab:models}. The
horizontal axis shows the ratio of the $\alpha$ sensitivity limit
including correlations to the one from statistics and systematics only
(which corresponds to $1$), where we only include fit models with
relevant model parameter contributions. Two models are highly
correlated if a possible signature in one model can be compensated by
a change of parameter(s) in the other.  Since we include the standard
neutrino oscillation scenario in all models, a small change in the fit
neutrino oscillation parameters might also compensate a damping
signature within the measurement precision of the
experiment. Therefore, we include for all signatures the standard
neutrino oscillation parameter correlation as dark bars, \ie, the dark
bars represent the fit against the standard neutrino oscillation
scenario (for fixed fit parameter $\alpha$), and the light bars are a
measure for the additional problem to distinguish a non-standard
signature from the ones of other possible non-standard models. The
interpretation of these bars is as follows: The dark bars reflect the
limit (right edges) for $\alpha$ (as multiple of the statistics limit)
beyond which the non-standard signature could be distinguished from
the standard neutrino oscillation case at the $3 \sigma$ confidence
level. However, if $\alpha$ should be within one of the light bar
ranges, then it could not be uniquely identified, since it could also
well be the non-standard signature corresponding to this bar. {}From
\figu{corrplot}, we make a number of interesting observations:
\begin{itemize}
\item
 Signatures which have negative $\gamma$'s (Absorption and Quantum
 decoherence I) are almost not affected by correlations with the
 neutrino oscillation parameters, \ie, they cannot be explained by
 different neutrino oscillation parameter values. In these cases, the
 spectrum is more suppressed for large values of $E$ than for small
 values, which means that the signature behaves unlike an oscillation
 signature corresponding to $\gamma = 2$. However, it is difficult to
 identify which of these models is realized.
\item
 Signatures with $\gamma=2$ (Oscillations into $\nu_s$ and Quantum
 decoherence II) are highly affected by correlations with the standard
 neutrino oscillation parameters, since the signatures have an energy
 dependence similar to the oscillation signature. Similar signatures,
 such as decay, can enhance this correlation.
\item
 Unique signatures (Wave packet decoherence and Neutrino decay) can
 easily be distinguished from all the other models. Although there could be
 some correlations with similar signatures for neutrino decay, the
 absolute impact on the $\alpha$ sensitivity limit is comparatively
 small (up to a factor of three).
\end{itemize}

\section{Summary and conclusions}
\label{sec:summary}

We have introduced exponential damping factors in the neutrino
oscillation probabilities, which lead to distinctive signatures, \ie,
energy dependent damping effects in the energy spectrum. These damping
factors are one approach to test non-oscillation effects on
the neutrino oscillation probability level. They can be motivated by
many different models such as intrinsic wave packet decoherence,
neutrino decay, oscillations into sterile neutrinos, neutrino
absorption, quantum decoherence, \etc. They describe the second order
contributions of small possible ``non-standard'' corrections to the
three-flavor neutrino oscillation framework (in vacuum as well as in
matter) on a rather abstract level. As opposed to tests of probability
conservation, the damping factors can, in addition, describe a damping
of the oscillating terms (which preserves the total probability) as
well as they imply, by their energy dependence, some information on
the type of effect.  We have demonstrated how damping factors can
modify the neutrino oscillation probabilities relevant for future
high-precision short- and long-baseline experiments, since these
experiments might be most sensitive to very small spectral effects.

As one application, we have shown that decoherence-like damping
signatures can severely modify the interpretation of experiments,
where we have chosen wave packet decoherence damping at new
short-baseline reactor experiments as an example. In this case, two
competing small effects, namely the effect of a non-zero value of
$\stheta$ and a damping contribution, might be mixed up.  In
particular, the damping could fake a value of $\stheta$ which is much
smaller than the value provided by Nature. Such a $\stheta$
suppression effect can either be intrinsic (such as quantum
decoherence), experiment dependent (such as some averaging effect not
taken into account), or both (such as wave packet decoherence related
to the production and detection processes). Intrinsic effects will be
observable by all types of experiments, which means that there are
very stringent limits available from existing data as well as future
experiments will test the consistency of the picture. On the other
hand, experiment dependent effects can only be checked by
complementary techniques measuring the same quantity.  One such
complementary pair has, in the past, been the solar and long-baseline
reactor experiments. In the future, it will therefore be very
important to measure $\stheta$ by reactor experiments and superbeams
as complementary techniques, since one of them alone could fail for
such experiment dependent effects. Eventually, the LSND experiment
could be a strong hint for such an experiment dependent effect if it
is rejected by the MiniBooNE experiment.

One of the most interesting features of damping signatures are their
characteristic spectral (energy) dependencies, which can act as a
``fingerprint'' for many sources of non-oscillation effects. For
example, specific signatures could point to new interesting physics
beyond the standard model. We have therefore discussed how large the
effects from different damping signatures have to be in order to be
identified and how well these damping signatures could be
distinguished for the example of neutrino factories.  In some cases,
such damping signatures can be compensated by a shift of the neutrino
oscillation parameters, which means that given such a damping effect,
it is quite likely to obtain an erroneous determination of these
parameters. However, if the damping effects are strong enough, then an
establishment of non-oscillation effects will be possible. Once
such a damping effect is established, it will be very interesting to
know from which non-standard mechanism it actually arises. Given this
question of the identification problem, we have found that signatures
with a damping similar to $\exp ( - \alpha L^\beta/ E^\gamma)$,
$\gamma = 1,2,\hdots$ are strongly correlated (peaking at $\gamma=2$)
with the standard neutrino oscillation parameters, \ie, it is
difficult to distinguish them from small adjustments in the neutrino
oscillation parameters. However, damping signatures similar to $\exp (
- \alpha L^\beta E^2 )$ can be very easily disentangled from the
neutrino oscillation parameters, but it is difficult to distinguish
them from each other. It is also extremely difficult to establish a
damping of the oscillations against a damping of the probabilities
with the same spectral index $\gamma$ because of the correlations with
the neutrino oscillation parameters.

Finally, we conclude that spectral tests of damping signatures in
neutrino oscillation probabilities are an important test of the
consistency of the three-flavor neutrino oscillation picture. If any
deviation from this picture is found, then the most important question
will be what sort of effect we are dealing with. Exactly this
information could be provided by the spectral dependence of the
damping signature, which means that this approach could be an
important test of physics beyond the standard model.

\subsection*{Acknowledgments}

We would like to thank John Bahcall, Manfred Lindner, and Thomas
Schwetz for useful discussions.

T.O. and W.W. would like to thank the IAS and the KTH respectively for the
warm hospitality and the financial support during their respective research
visits.

This work was supported by the Royal Swedish Academy of Sciences
(KVA), the Swedish Research Council (Vetenskapsr{\aa}det), Contract
Nos.~621-2001-1611, 621-2002-3577, the G{\"o}ran Gustafsson
Foundation, the Magnus Bergvall Foundation, the W.~M.~Keck
Foundation, and NSF grant PHY-0070928.

\providecommand{\bysame}{\leavevmode\hbox to3em{\hrulefill}\thinspace}

\end{document}